\documentclass[twocolumn]{aastex631}
\usepackage[T1]{fontenc}

\usepackage{amsmath,graphicx,hyperref}
\usepackage{natbib,threeparttable,rotating}
\usepackage{ulem}
\usepackage{multirow}
\newcommand{\etal}{et al.}

\begin{document}

\title{The Cosmic Ultraviolet Baryon Survey (CUBS) VIII: Group Environment of the Most Luminous Quasars at $z\approx1$} 

\author[0000-0002-0311-2812]{Jennifer~I-Hsiu~Li}
\affiliation{Department of Astronomy, University of Michigan, Ann Arbor, MI 48109, USA}
\affiliation{Michigan Institute for Data Science, University of Michigan, Ann Arbor, MI 48109, USA}

\author[0000-0001-9487-8583]{Sean~D.~Johnson}
\affiliation{Department of Astronomy, University of Michigan, Ann Arbor, MI 48109, USA}

\author[0000-0003-3244-0409]{Erin~Boettcher}
\affiliation{Department of Astronomy, University of Maryland, College Park, MD 20742, USA}
\affiliation{X-ray Astrophysics Laboratory, NASA/GSFC, Greenbelt, MD 20771, USA}
\affiliation{Center for Research and Exploration in Space Science and Technology, NASA/GSFC, Greenbelt, MD 20771, USA}

\author[0000-0001-5804-1428]{Sebastiano~Cantalupo}
\affiliation{Department of Physics, University of Milan Bicocca, Piazza della Scienza 3, I-20126 Milano, Italy}

\author[0000-0001-8813-4182]{Hsiao-Wen~Chen}
\affiliation{Department of Astronomy and Astrophysics, The University of Chicago, Chicago, IL 60637, USA}

\author[0000-0002-8739-3163]{Mandy~C.~Chen}
\affiliation{Department of Astronomy and Astrophysics, The University of Chicago, Chicago, IL 60637, USA}

\author{David~R.~DePalma}
\affiliation{MIT‚ÄìKavli Institute for Astrophysics and Space Research, 77 Massachusetts Avenue, Cambridge, MA 02139, USA}

\author[0000-0002-2662-9363]{Zhuoqi (Will) Liu}
\affiliation{Department of Astronomy, University of Michigan, Ann Arbor, MI 48109, USA}

\author[0000-0002-9141-9792]{Nishant Mishra}
\affiliation{Department of Astronomy, University of Michigan, Ann Arbor, MI 48109, USA}

\author{Patrick~Petitjean}
\affiliation{Institut d'Astrophysique de Paris
98bis Boulevard Arago, 75014, Paris, France}

\author[0000-0002-2941-646X]{Zhijie~Qu}
\affiliation{Department of Astronomy and Astrophysics, The University of Chicago, Chicago, IL 60637, USA}

\author[0000-0002-8459-5413]{Gwen~C.~Rudie}
\affiliation{The Observatories of the Carnegie Institution for Science, 813 Santa Barbara Street, Pasadena, CA 91101, USA}

\author[0000-0002-0668-5560]{Joop~Schaye}
\affiliation{Leiden Observatory, Leiden University, PO Box 9513, NL-2300 RA Leiden, the Netherlands}

\author[0000-0001-7869-2551]{Fakhri~S.~Zahedy}
\affiliation{The Observatories of the Carnegie Institution for Science, 813 Santa Barbara Street, Pasadena, CA 91101, USA}

\shorttitle{CUBS VIII: Quasar environment}
\shortauthors{Li \etal}

\begin{abstract}
We investigate the group-scale environment of 15 luminous quasars (luminosity $L_{\rm 3000}>10^{46}$\,erg\,s$^{-1}$) from the Cosmic Ultraviolet Baryon Survey (CUBS) at redshift $z\approx1$. Using the Multi Unit Spectroscopic Explorer (MUSE) integral field spectrograph on the Very Large Telescope (VLT), we conduct a deep galaxy redshift survey in the CUBS quasar fields to identify group members and measure the physical properties of individual galaxies and galaxy groups. We find that the CUBS quasars reside in diverse environments. The majority (11 out of 15) of the CUBS quasars reside in overdense environments with typical halo masses exceeding $10^{13}{\rm M}_{\odot}$, while the remaining quasars reside in moderate-size galaxy groups. No correlation is observed between overdensity and redshift, black hole (BH) mass, or luminosity. 
Radio-loud quasars (5 out of 15 CUBS quasars) are more likely to be in overdense environments than their radio-quiet counterparts in the sample, consistent with the mean trends from previous statistical observations and clustering analyses. Nonetheless, we also observe radio-loud quasars in moderate groups and radio-quiet quasars in overdense environments, indicating a large scatter in the connection between radio properties and environment. We find that the most UV luminous quasars might be outliers in the stellar mass-to-halo mass relations or may represent departures from the standard single-epoch BH relations.
\end{abstract}

\section{Introduction}

In hierarchical structure formation models, small initial density fluctuations grow to form galaxies and galaxy clusters in the present Universe. 
Residing at the centers of massive galaxies, quasars can provide substantial feedback into the surrounding environment through radio jets and direct heating of the interstellar and circumgalactic media (ISM/CGM) \citep[e.g.,][]{Silk_Rees_1998, Heckman_Best_2014}. The observed tight correlations between SMBHs and host galaxy properties support the BH-galaxy co-evolution models \citep[e.g.,][and references therein]{Magorrian_etal_1998, Ferrarese_Merritt_2000, Gebhardt_etal_2000, Kormendy_Ho_2013}. Studying the environment quasars reside in provides insights into galaxy/BH evolution, the interplay between quasars, their surrounding gas reservoirs, and hierarchical structure formation.

\begin{table*}
\movetableright=-0.4in
\caption{Summary of the quasar properties}
\label{tab:quasar}
\begin{tabular}{llcccccr}
\hline\hline
 Quasar Name &   Other Name &       R. A. &          Decl. &      z & $\log M_{\rm BH}/{\rm M}_{\odot}$ & $\log L_{\rm 3000}/$erg$\,$s$^{-1}$ &    $R$ \\
 & & (J2000)  &  (J2000)    &  &  &  &  \\
\hline
J0028$-$3305 & & 00:28:30.41 & $-$33:05:49.25 & 0.8873 &                     9.37 &              46.3 & $<$11 \\
J0110$-$1648 &  & 01:10:35.51 & $-$16:48:27.70 & 0.7823 &                     9.12 &              46.2 &   230 \\
J0111$-$0316 &  & 01:11:39.17 & $-$03:16:10.89 & 1.2384 &                     9.95 &              46.9 &  $<$6 \\
J0114$-$4129 & HE0112$-$4145 & 01:14:22.12 & $-$41:29:47.29 & 1.0238 &                     9.45 &              46.2 &    18 \\
J0119$-$2010 &  & 01:19:56.09 & $-$20:10:22.73 & 0.8163 &                     9.32 &              46.4 &  $<$7 \\
J0154$-$0712 &  & 01:54:54.68 & $-$07:12:22.17 & 1.2930 &                     9.53 &              46.8 &  $<$8 \\
J0248$-$4048 & HE0246$-$4101 & 02:48:06.29 & $-$40:48:33.66 & 0.8844 &                     9.65 &              46.7 &     4 \\
J0333$-$4102 & HE0331$-$4112 & 03:33:07.08 & $-$41:02:01.15 & 1.1153 &                     9.88 &              46.8 &  $<$6 \\
J0357$-$4812 & PKS0355$-$483 & 03:57:21.92 & $-$48:12:15.16 & 1.0128 &                     9.64 &              46.3 &   783 \\
J0420$-$5650 & HE0419$-$5657 & 04:20:53.91 & $-$56:50:43.96 & 0.9481 &                     9.15 &              46.1 & $<$18 \\
J0454$-$6116 &  & 04:54:15.95 & $-$61:16:26.56 & 0.7861 &                     9.36 &              46.2 &     7 \\
J2135$-$5316 &  & 21:35:53.20 & $-$53:16:55.82 & 0.8115 &                     9.37 &              46.6 &  $<$4 \\
J2245$-$4931 & PKS2242$-$498 & 22:45:00.21 & $-$49:31:48.46 & 1.0011 &                     9.63 &              46.3 &  2779 \\
J2308$-$5258 & HE2305$-$5315 & 23:08:37.80 & $-$52:58:48.94 & 1.0733 &                     9.46 &              46.5 & $<$10 \\
J2339$-$5523 & HE2336$-$5540 & 23:39:13.22 & $-$55:23:50.84 & 1.3544 &                     9.98 &              47.1 &   339 \\
\hline\hline
\end{tabular}

\tablecomments{
The table columns are quasar name, alternative quasar names in the literature, quasar right ascension, declination, redshift, BH mass, luminosity at 3000\,${\rm \AA}$, and radio loudness ($R$).}
\end{table*}

Previous studies of quasars and their environments often relied on wide-field optical/infrared imaging surveys and clustering analyses. The overdensity of quasar environments is estimated by comparing the galaxy number counts within certain projected distances around known quasars and background fields. Many studies found that the galaxy number density around quasars is higher than around inactive galaxies of similar masses \citep[e.g.,][]{Serber_etal_2006, Wylezalek_etal_2013, Karhunen_etal_2014}. This could indicate that quasars reside in relatively massive halos since environmental density is tightly correlated with halo mass \citep{Haas_etal_2012}. Radio-loud quasars, in particular, are found to reside in denser environments compared to radio-quiet quasars and inactive galaxies of similar masses over a wide redshift range \citep[e.g.,][]{RamosAlmeida_etal_2013, Wylezalek_etal_2013, Hatch_etal_2014}. If radio-loudness is connected to the overdensity of quasar environments or quasar properties, then this could be evidence connecting merger and galaxy interaction to the triggering of nuclear activity and jet formation \citep[e.g., ][]{DiMatteo_etal_2005, Hopkins_etal_2006}. On the other hand, it is unclear if overdensity correlates with other quasar properties, e.g., redshift, quasar luminosity, and BH mass \citep[e.g.,][]{Serber_etal_2006, Karhunen_etal_2014}. These results are broadly consistent with clustering analyses at low redshift, where there is a weak luminosity dependency in quasar clustering and radio-loud quasars typically reside in more clustered environments than radio-quiet quasars at fixed optical luminosity \citep[e.g.,][]{Shen_etal_2009, Zhang_etal_2013, Shen_etal_2013}.

However, these methods can only provide a statistical view of the quasar environments, since the foreground or background galaxies in the fields cannot be removed without measuring individual galaxy redshifts. \cite{Stott_etal_2020} 
found that 8 out of 12 quasar fields display galaxy overdensities at $1<z<2$ using a deep HST grism spectroscopy survey. However, \cite{Wethers_etal_2022} found that quasars and galaxies reside in similar-sized galaxy groups when controlled for stellar mass and redshift, arguing against the scenario of merger/interaction triggered quasar activity. Instead, they find quasars are more likely to be central galaxies, which potentially links quasar activities to either gas accretion or rich group environments. In addition, \cite{Stone_etal_2023} found that the galaxy group members around quasars and their inactive counterparts have similar stellar properties, suggesting that quasar feedback does not have a strong influence on the galaxy group members. With these mixed results, more comprehensive studies of individual quasars, their group environment and neighbors, and surrounding gas flow are needed to understand the interplay between quasars and their environments. 

Wide-field integral field spectrographs, like the Multi Unit Spectroscopic Explorer (MUSE) on the Very Large Telescope \citep[VLT,][]{MUSE}, are capable of capturing the galaxy group environment by providing redshifts for all galaxies in the field of view (above a certain flux threshold). 
In this paper, we characterize the environment of 15 quasars observed by MUSE in the Cosmic Ultraviolet Baryon Survey \citep[CUBS,][]{Chen_etal_2020_CUBS1}, and how they depend on the central BH properties. The 15 CUBS quasars are selected from the brightest quasars in the GALEX near-UV bandpass (NUV; 1770-2730\,\AA) at $z\approx1$ in the Dark Energy Survey footprint \citep{DES_Y3}. The quasar fields are selected without prior knowledge of the surrounding galactic environment or radio properties. 

This paper is organized as follows. We describe the CUBS survey and relevant follow-up observations in Section \ref{sec:obs} and our data analysis in Section \ref{sec:analysis}. The main results are presented in Section \ref{sec:results}. We discuss our results in Section \ref{sec:discussion} and conclude in Section \ref{sec:conclusions}. Throughout this paper, we adopt a flat $\Lambda$CDM cosmology with $\Omega_M=0.3$, $\Omega_{\lambda}=0.7$, and $H_0=70\,{\rm km\,s^{-1}\,Mpc^{-1}}$.

\begin{figure}
\centering
    \includegraphics[width=0.45\textwidth]{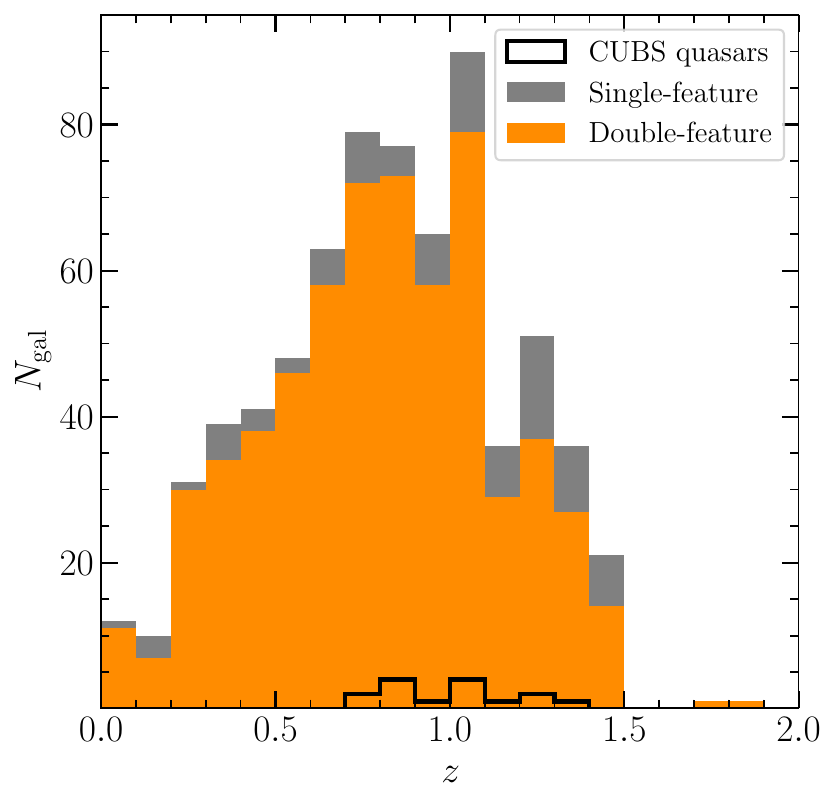}
\caption{Redshift distribution for the CUBS quasars (black), and the double-feature (orange) and single-feature (grey) galaxy redshifts identified in 15 quasar fields.}
\label{fig:redshift}
\end{figure}

\section{Observations}\label{sec:obs}

The main goal of CUBS is to map the cosmic baryonic reservoir at intermediate redshift ($0.8\lesssim\,z\lesssim\,1.4$) using high-quality UV absorption spectroscopy and deep ground-based optical and near-infrared observations. The UV absorption spectroscopy is obtained from the Cosmic Origins Spectrograph \citep[COS, ][]{COS} in a large Hubble Space Telescope (HST) Cycle 25 General Observer Program (GO-CUBS; PID = 15163; PI: Chen). The ground-based spectroscopic observations consist of three main components using the Inamori-Magellan Areal Camera and Spectrograph \citep[IMACS;][]{IMACS} on the Magellan Baade Telescope, the Low Dispersion Survey Spectrograph 3 (LDSS3) on the Magellan Clay Telescope, and the Multi Unit Spectroscopic Explorer \citep[MUSE;][]{MUSE} on the Very Large Telescope (VLT) that cover different depths and projected angular radius from the quasar sight lines. All quasar fields are covered by the Dark Energy Survey footprint, and additional $H$-band photometry is obtained by the Four Star Infrared Camera \citep{Fourstar} on the Magellan Telescopes. The detailed survey design is described in \cite{Chen_etal_2020_CUBS1}, here we summarize the observations and data analysis relevant to this work.

\subsection{MUSE Observation}

The MUSE galaxy survey is the deepest and most compact component of our ground-based optical spectroscopic observations, aiming to target galaxies as faint as 0.01\,$L_{*}$ at $z=1$ within $\lesssim30''$ from the central quasar. The MUSE survey is carried out on the VLT UT4 in service mode under program ID 0104.A-0147 (PI: Chen). The observations cover $1'\times\,1'$ fields of view with plate scales of $0.2''$ and spectral resolution of  $\approx120$\,km\,s$^{-1}$ at 7000\,\AA. The observations are in the wide-field mode (WFM) with ground layer adaptive optics (AO) assistance to ensure uniform image quality of $<0.8''$ across all fields. {The limiting magnitude (3$\sigma$) of the detected sources in the MUSE datacubes are ${AB}(r)\approx26$ in the psuedo-$r$ band \citep{Qu_etal_2023}}, sufficient to identify faint galaxies down to ${\rm M}_{\rm \star} \approx 10^{8}-10^{9} M_{\odot}$.
The MUSE data cubes are reduced using the standard ESO MUSE pipeline \citep{Weilbacher_etal_2020} and the custom data reduction package {\tt CUBEXTRACTOR} \citep{Cantalupo_etal_2019}. Additional details of the MUSE observations are described in \cite{Chen_etal_2020_CUBS1}.

\subsection{FourStar H-Band Imaging}
Deep $H$-band images were obtained in October 2017 using the FourStar Infrared Camera \citep{Fourstar} on the Magellan Telescopes. The data reduction was performed using the FourCLift custom package \citep[details described in][]{Kelson_etal_2014} and following the procedures in \cite{Kelson_etal_2014} and \cite{ Rudie_etal_2017}.
The exposure in each field is roughly 2000$-$3000 seconds, {and the limiting magnitude (3$\sigma$) of the detected sources are $\approx24.5$ mag \citep{Qu_etal_2023}.} The median seeing of the final images is $\approx$0.5$''$. 

\begin{figure}
\centering
    \includegraphics[width=0.45\textwidth]{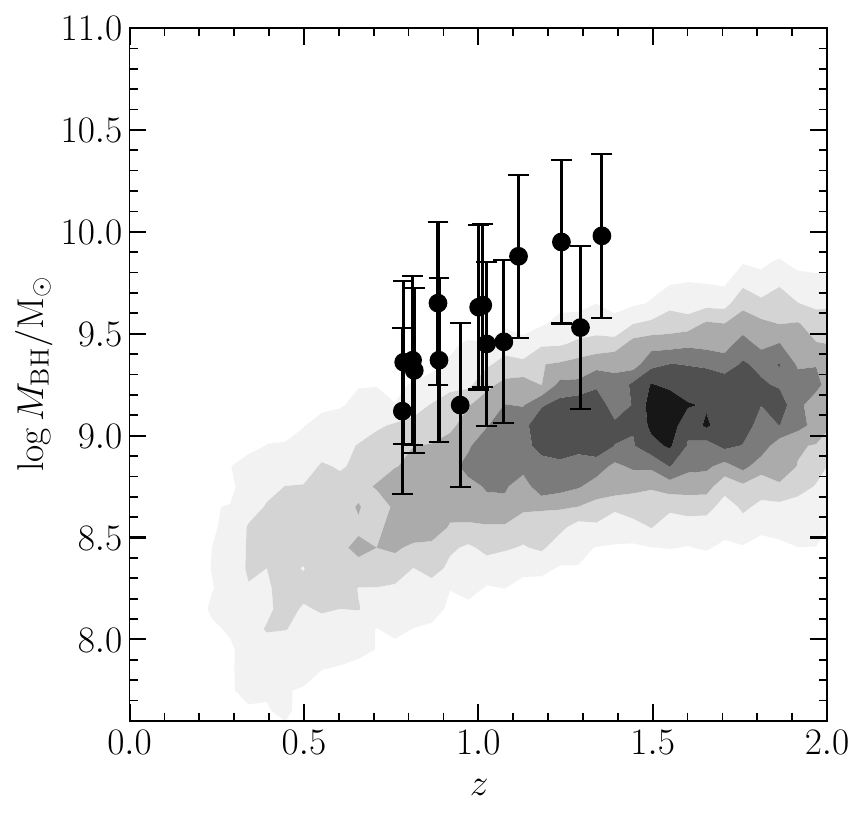}
    \includegraphics[width=0.45\textwidth]{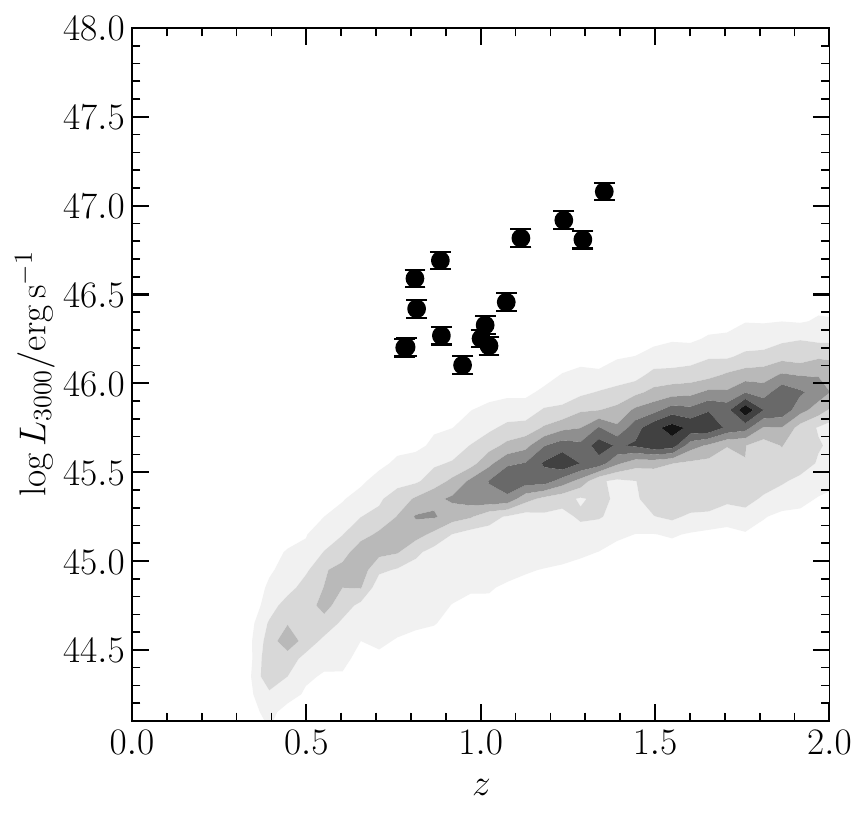}
\caption{BH mass (top), luminosity at 3000\,${\rm \AA}$ (bottom), and redshift distribution of the CUBS quasars. The shaded contours show the quasar properties from the DR7 quasar catalog, containing quasars brighter than $M_{i}=-22$ with a redshift range of $0<z<5$ \citep{Shen_etal_2011_sdssq}. The error bars include measurement uncertainties from spectral fitting and systematic uncertainties (0.4\,dex for BH mass and 0.05\,dex for luminosity). The CUBS quasars are roughly $\approx0.5$\,dex and $\approx1$\,dex higher than the general quasar population in BH mass and luminosity, respectively.}
\label{fig:quasar}
\end{figure}

\subsection{Supplementary Spectra}
While using the MUSE WFM AO observation setup ensures a uniform imaging resolution, the use of wavefront lasers produces a gap at 5800 to 5965\,\AA\ in the spectra. We obtained supplementary spectra from the Magellan Echellette (MagE) Spectrograph for quasars with the MgII line in/near the sodium gap (J0333, J2135, J2245, J2308) to ensure good spectral coverage around the broad MgII line for BH mass estimation. The observations were carried out on September 29 and 30 2021 on the Magellan Baade Telescope. Two to three exposures of 300--600 seconds, depending on the quasar luminosity, were taken for each source to mitigate cosmic rays. The final reduced spectral range is 4000\,\AA\ to 9900\,\AA\ with median spectral resolution of $\approx$1\,\AA\, and spectral sensitivity ($1\sigma$) of 1--4$\times10^{-17}$ erg\,cm$^{-2}$\,s$^{-1}$\,\AA$^{-1}$. We follow the standard data reduction pipeline with {\tt CarPy} \citep{Carpy1,Carpy2} to reduce the data and match the flux scale to the MUSE spectra. 

\begin{figure*}
\raggedright
    \includegraphics[width=0.24\textwidth]{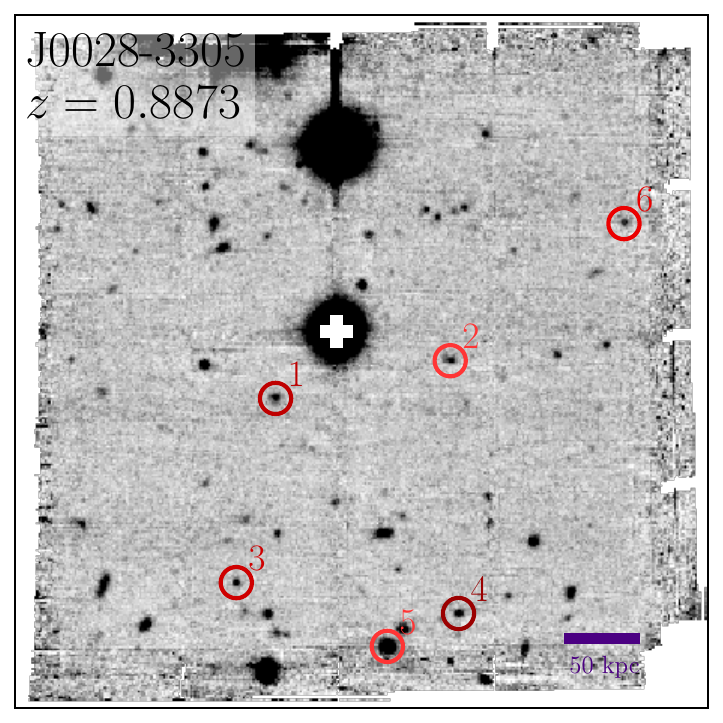}
    \includegraphics[width=0.24\textwidth]{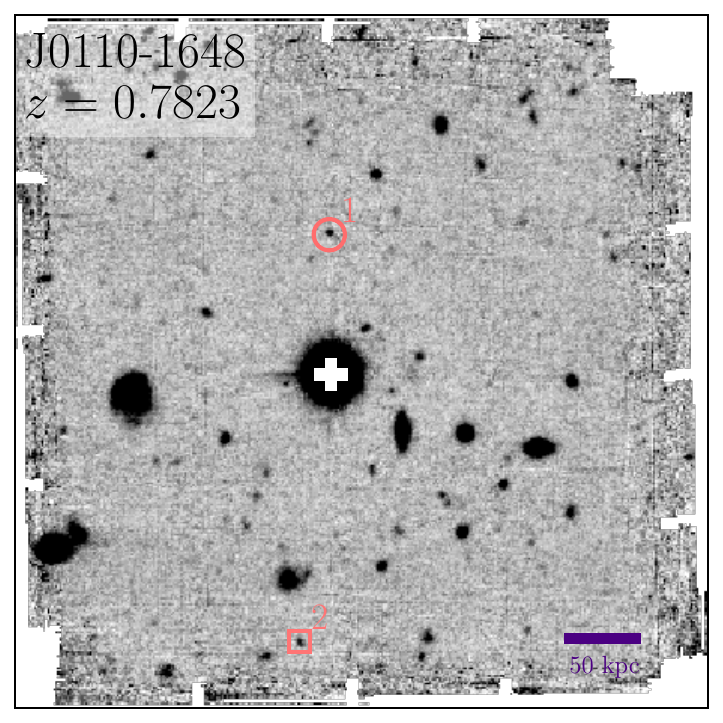}
    \includegraphics[width=0.24\textwidth]{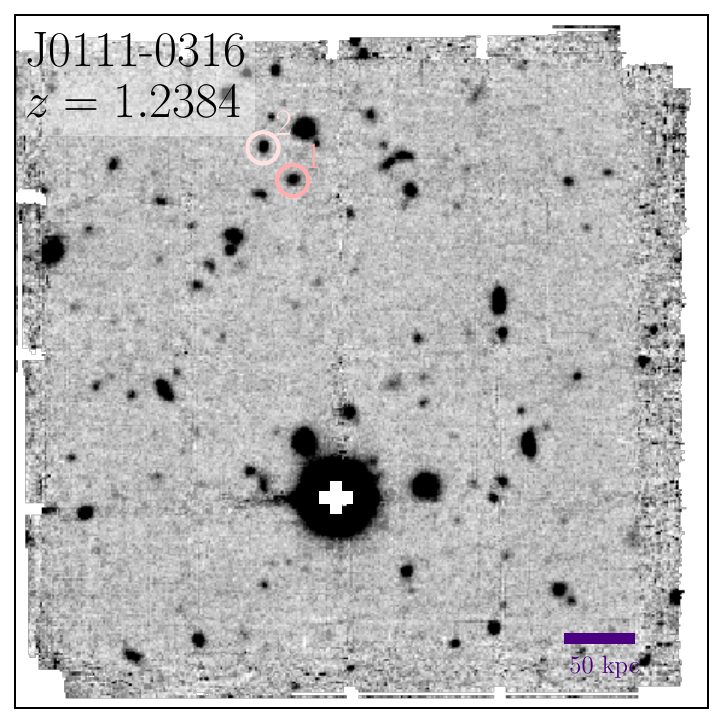} 
    \includegraphics[width=0.24\textwidth]{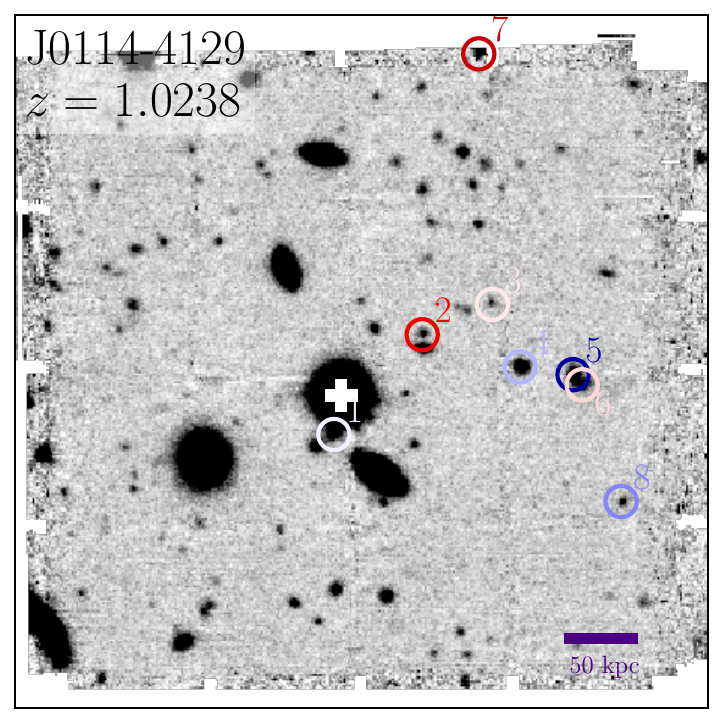} \\
    \includegraphics[width=0.24\textwidth]{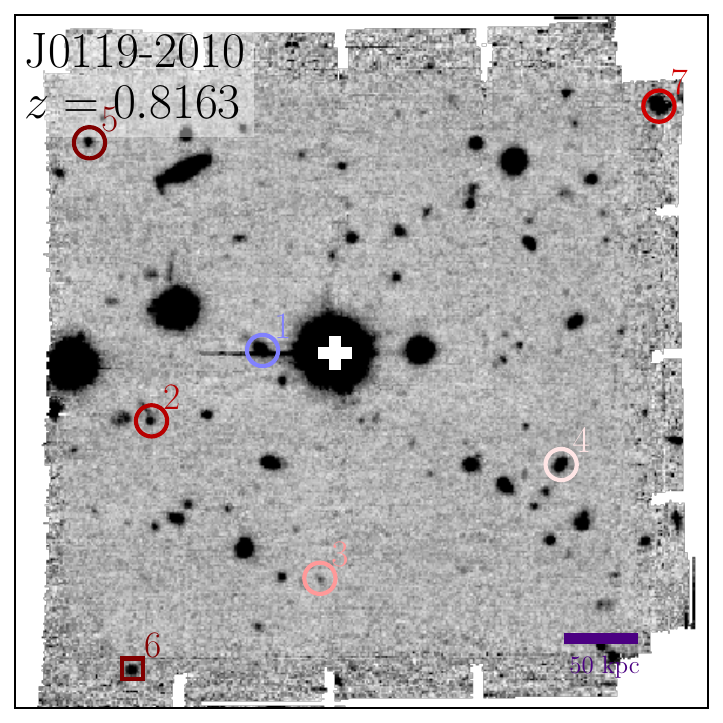}
    \includegraphics[width=0.24\textwidth]{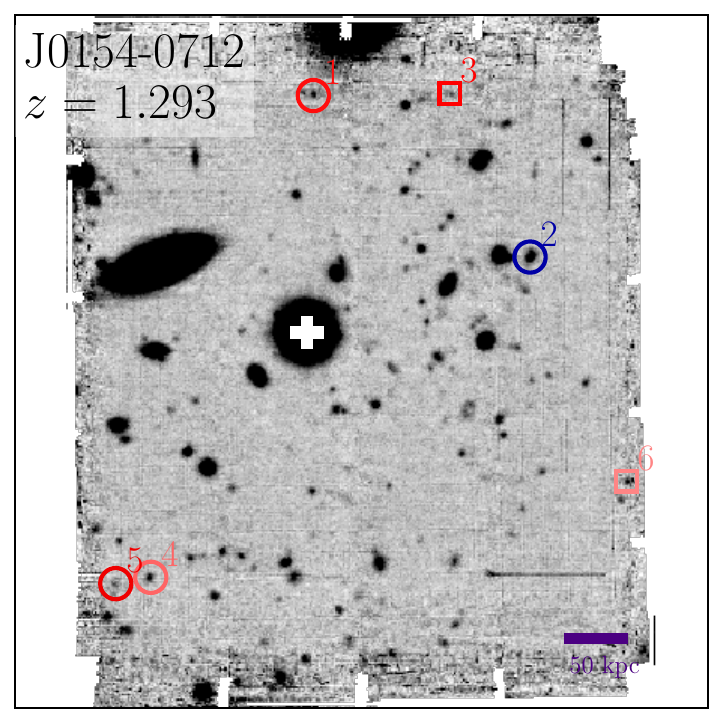} 
    \includegraphics[width=0.24\textwidth]{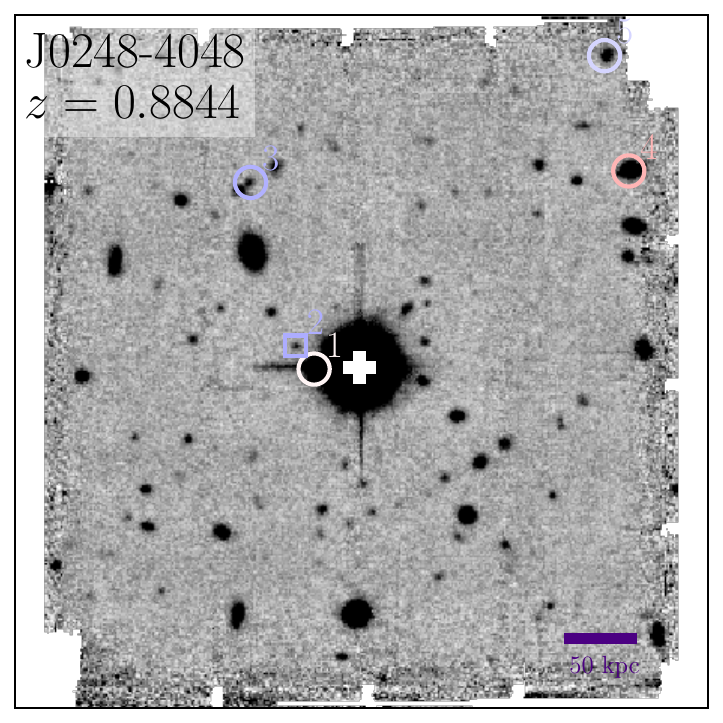}
    \includegraphics[width=0.24\textwidth]{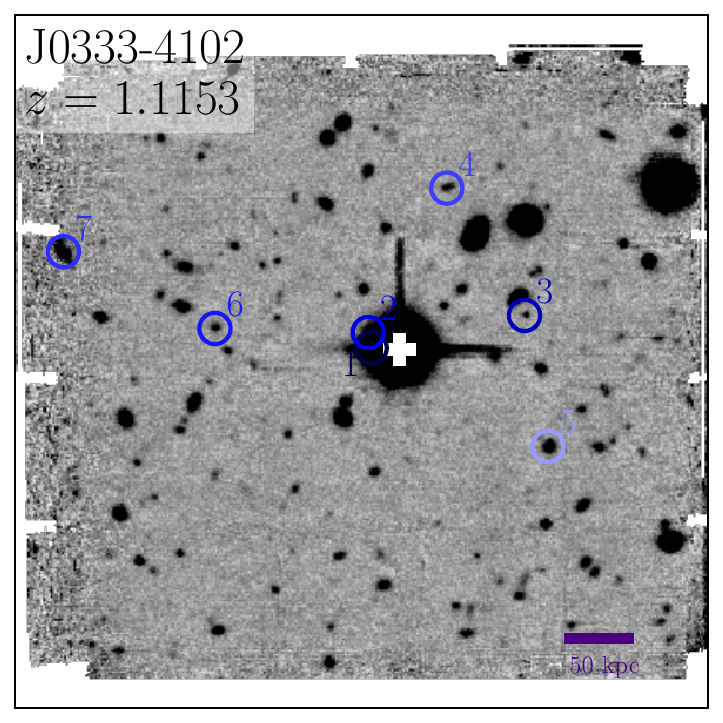} \\
    \includegraphics[width=0.24\textwidth]{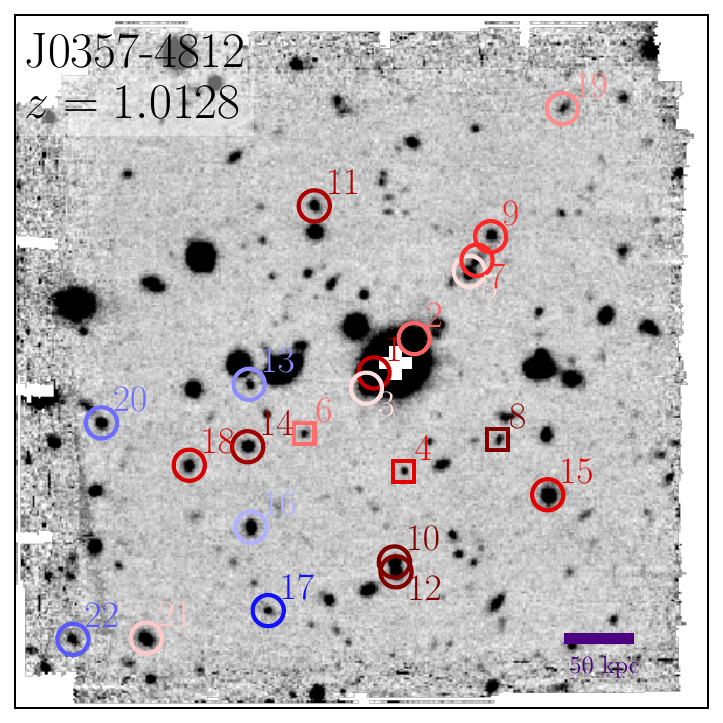} 
    \includegraphics[width=0.24\textwidth]{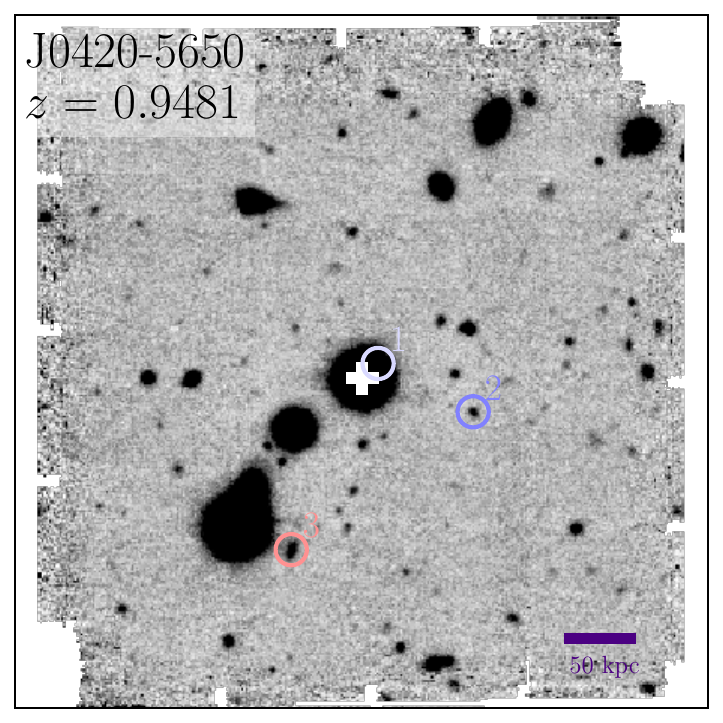}
    \includegraphics[width=0.24\textwidth]{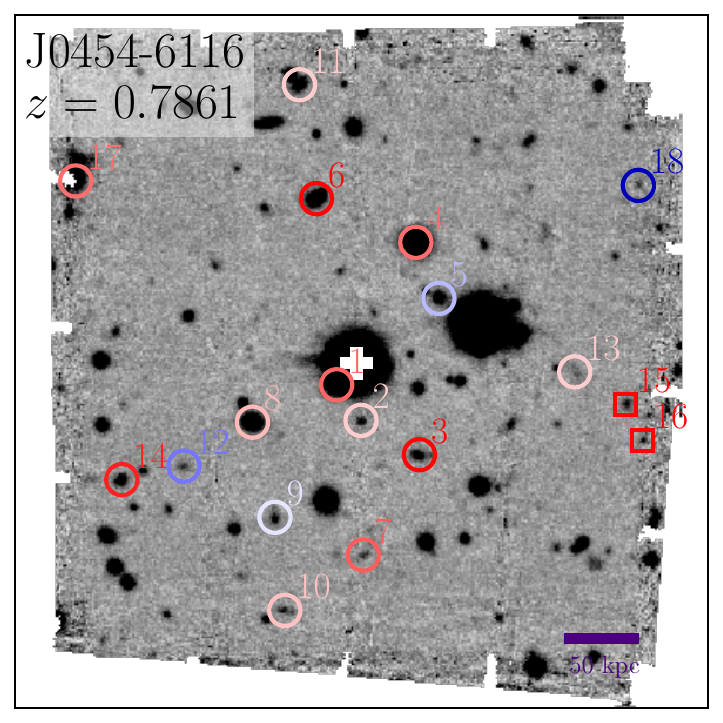}
    \includegraphics[width=0.24\textwidth]{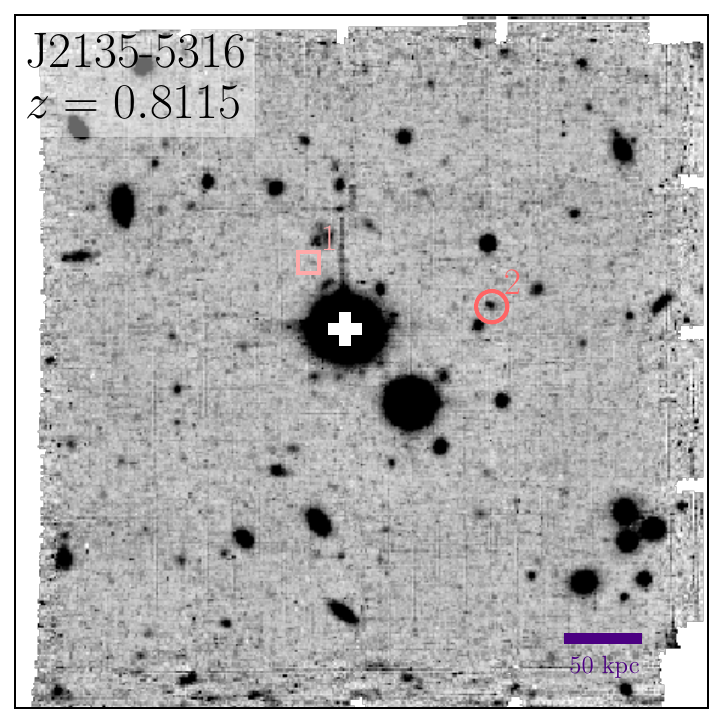} \\
    \includegraphics[width=0.24\textwidth]{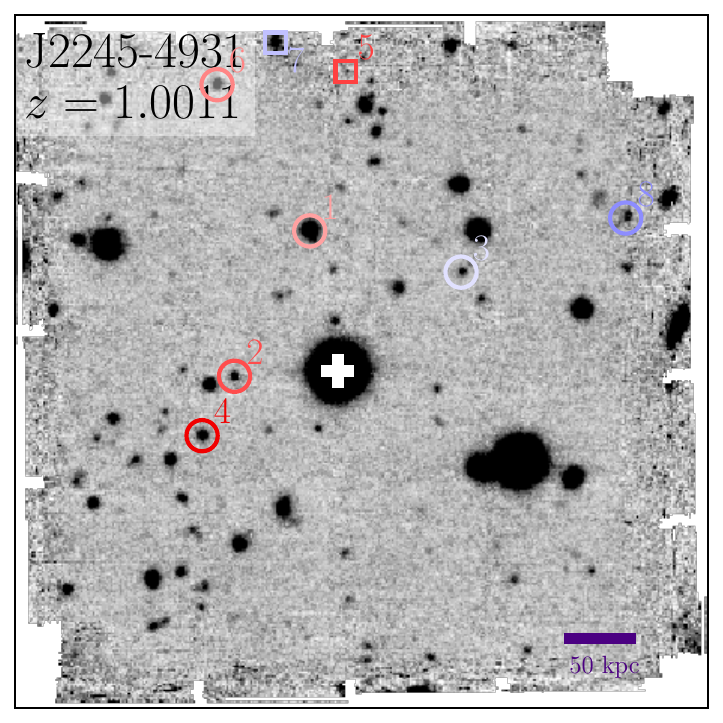}
    \includegraphics[width=0.24\textwidth]{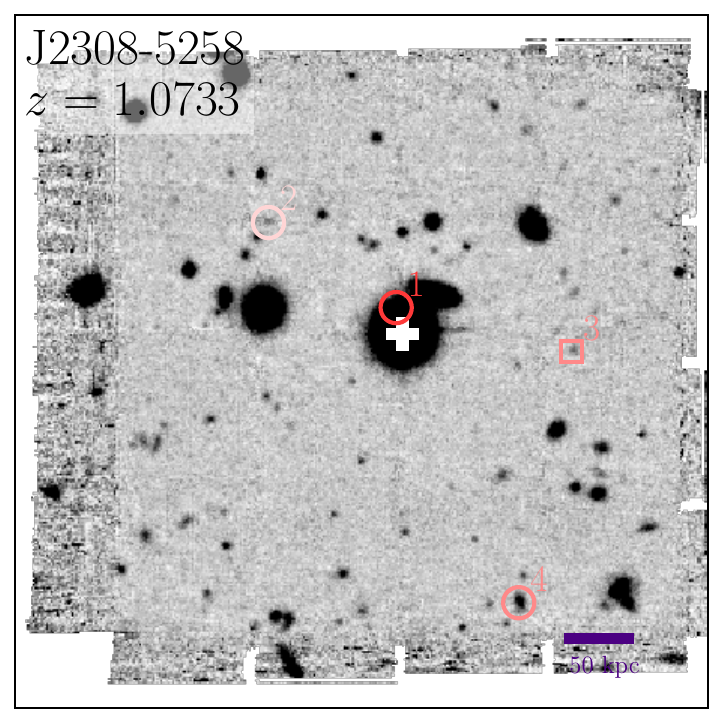}
    \includegraphics[width=0.24\textwidth]{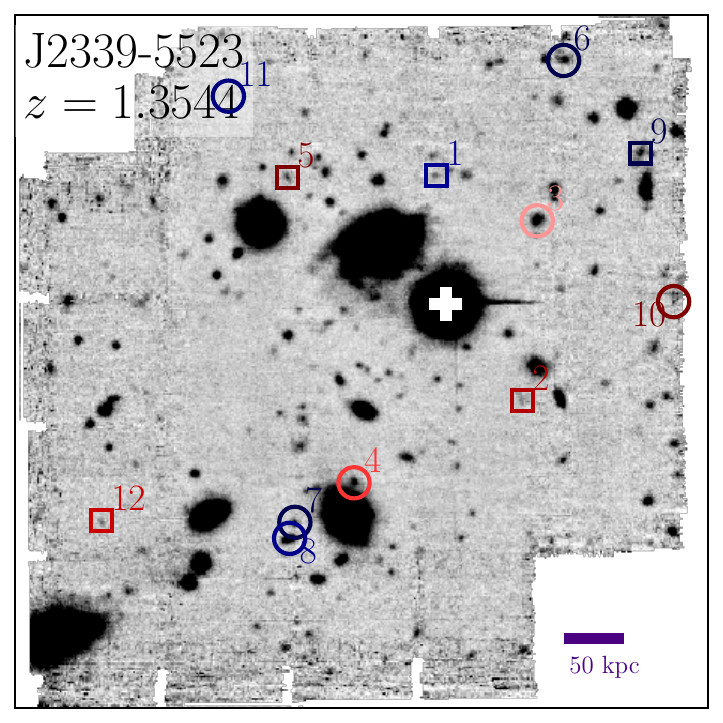} 
    \includegraphics[height=0.24\textwidth]{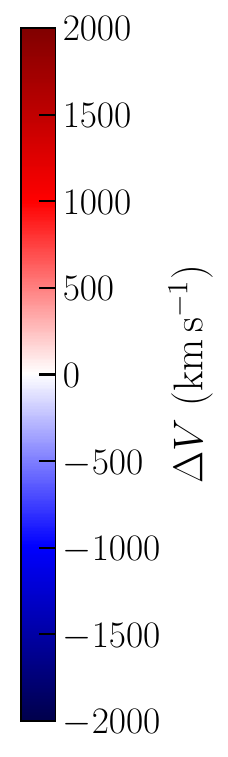} \\
\caption{The white-light images of each quasar field constructed from the MUSE data. The CUBS quasars are located at the plus signs, and the identified quasar group galaxies are labeled (the circles show the galaxies with double-feature redshifts, and the squares show the galaxies with single-feature redshifts) and colored by the relative velocity from the quasar systematic redshift. The horizontal bar on the lower right corner of each panel indicates a physical scale of 50\,kpc.}
\label{fig:field}
\end{figure*}

\section{Data Analysis}\label{sec:analysis}
\subsection{MUSE Redshift Survey}
We identify all continuum sources in the MUSE white-light image with {\tt Source Extractor} \citep{SourceExtractor} for redshift measurements. For the quasars, we extract the optical spectra within apertures that include $>$95\% of the quasar light and measure the redshifts based on narrow [O II] emission lines, which results in a typical redshift uncertainty of 20--30\,km\,s$^{-1}$ \citep{Hewett_Wild_2010}. For galaxies, the spectra of each source are extracted within a $\approx$0.6\,$''$ radius using the Python MUSE data analysis package {\tt mpdaf} \citep{Mpdaf}. We then follow the procedures described in \cite{Johnson_etal_2018} and \cite{Helton_etal_2021} to determine the redshift of each source. To briefly summarize, each spectrum is fitted with a linear combination of SDSS galaxy eigenspectra over a wide redshift grid, and the best-fit redshift is identified at the global minimum in the resulting $\chi^{2}$ grid. We then visually inspect the best-fit spectra to assign a quality ranking for each redshift. We define the robust ``double-feature'' redshifts to have good fits around at least two spectral features (including both emission and absorption, visually evaluated by trained team members) and the ``single-feature'' redshift to have good fits around one spectral feature, and the rest of the redshifts are discarded.
It is possible that the single-feature redshifts are not unique solutions, in these cases, we only include them in the final redshift measurement after all other possible solutions (e.g., a different emission line is well-fitted to the spectral feature) are ruled out. 

Figure \ref{fig:redshift} shows the redshift distribution. Across all 15 quasar fields, we identified 1542 continuum sources (excluding stars) and measured 712 ``double-feature'' redshifts and 109 ``single-feature'' redshifts. The redshift success rates are approximately 40--65\% across all fields. The primary reason for not measuring a redshift is usually because of low signal-to-noise ratios in the spectra or the lack of spectral features. Our MUSE redshift survey is most sensitive to redshifts $z<1.5$, where strong line features (e.g., H$\alpha$, H$\beta$, OIII, [OII] lines, and stellar absorption features) are within the MUSE spectral coverage. 

\begin{table*}
\movetableright=0.25in
\caption{Summary of the galaxy groups (within the MUSE field of view)}
\label{tab:group}
\begin{tabular}{lrrrrrrr}
\hline\hline
 Quasar Name & $N_{\rm group}$ &       $\delta_{\rm g}$ & $\sigma_{\rm \delta}$ & $\mu_{{\rm Vel}}$ & $\sigma_{{\rm Vel}}$ & $\sigma_{{\rm Vel, corr}}$ & $\log M_{\rm 200}/{\rm M}_{\odot}$ \\

&  &  &  & (km\,s$^{-1}$) & (km\,s$^{-1}$) & (km\,s$^{-1}$) & \\
\hline
J0028$-$3305 &             7 & $10.1_{-3.7}^{+4.8}$ &                2.73 &                  526 &                270 &                      280 &                      13.2 \\
J0110$-$1648 &             3 &  $3.7_{-2.3}^{+3.3}$ &                1.65 &                   -- &                -- &                      -- &                      -- \\
J0111$-$0316 &             3 &  $3.7_{-2.3}^{+3.3}$ &                1.65 &                   -- &                 -- &                       -- &                      -- \\
J0114$-$4129 &             9 & $13.2_{-4.2}^{+5.3}$ &                3.12 &                   20 &                400 &                      410 &                      13.6 \\
J0119$-$2010 &             8 & $11.6_{-4.0}^{+5.0}$ &                2.93 &                  437 &                460 &                      480 &                      13.9 \\
J0154$-$0712 &             7 & $10.1_{-3.7}^{+4.8}$ &                2.73 &                  189 &                420 &                      440 &                      13.7 \\
J0248$-$4048 &             6 &  $8.5_{-3.4}^{+4.5}$ &                2.50 &                $-$38 &                110 &                      110 &                      12.1 \\
J0333$-$4102 &             8 & $11.6_{-4.0}^{+5.0}$ &                2.93 &               $-$468 &                300 &                      310 &                      13.3 \\
J0357$-$4812 &            23 & $35.3_{-7.0}^{+8.1}$ &                5.01 &                  402 &                570 &                      580 &                      14.1 \\
J0420$-$5650 &             4 &  $5.3_{-2.7}^{+3.8}$ &                1.98 &                   -- &                -- &                      -- &                      -- \\
J0454$-$6116 &            19 & $29.0_{-6.4}^{+7.4}$ &                4.56 &                  157 &                290 &                      300 &                      13.3 \\
J2135$-$5316 &             3 &  $3.7_{-2.3}^{+3.3}$ &                1.65 &                   -- &                -- &                      -- &                      -- \\
J2245$-$4931 &             9 & $13.2_{-4.2}^{+5.3}$ &                3.12 &                  143 &                240 &                      250 &                      13.1 \\
J2308$-$5258 &             5 &  $6.9_{-3.1}^{+4.1}$ &                2.26 &                   -- &                -- &                      -- &                      -- \\
J2339$-$5523 &            13 & $19.5_{-5.2}^{+6.2}$ &                3.77 &               $-$216 &                940 &                      960 &                      14.6 \\
J2339$-$5523$^{a}$ &            7 & $10.1_{-3.7}^{+4.8}$ &  2.73 &               592 &               370 &                     390 &                      13.5 \\
\hline\hline
\end{tabular}
 \tablecomments{The table columns are quasar name, number of group galaxy members (including the quasar host galaxy), overdensity, significance of overdensity ($\sigma_{\delta}=\delta_{\rm g}/\Delta\delta_{g}$), mean velocity of the galaxy group with respect to the quasar, velocity dispersion of the galaxy group, the corrected velocity dispersion (see Section \ref{sec:discussion}), and the estimated halo mass. $^{a}$ The group measurements of J2339 when assuming a bimodal velocity distribution, the group with mean velocity closer to the quasar velocity is reported here.}
\end{table*}

\subsection{BH Masses}
BH masses are estimated through the single-epoch method \citep{Shen_etal_2011_sdssq}. Assuming the broad-line region (BLR) is virialized, we can measure the BH mass from a single spectrum by estimating the BLR size from the quasar luminosity with the radius-luminosity relation \citep{Bentz_etal_2013} and the BLR virial velocity from the width of a broad emission line (e.g., the 2800\,\AA\ MgII line). We extract quasar spectra from apertures of 10 pixels centered on the quasar positions in the MUSE data cube, which roughly encompass 95\% of the total light. For the quasars where the MgII line falls into the sodium gap, we use the supplementary quasar spectra from the MagE observations. We fit the quasar spectra with a power-law continuum, the FeII pseudo-continuum, and a series of Gaussian profiles for each broad line using the code {\tt PyQSOFit} \citep{Pyqsofit}. The host galaxy light is expected to be very faint compared to the bright core for these quasars, so we do not perform host decomposition in {\tt PyQSOFit}.
We include up to three broad components, one narrow component, and an additional wing component to model the MgII line. Finally, we follow the recipe from \cite{Shen_etal_2011_sdssq} to estimate BH masses using the luminosity at 3000\,\AA\ and the combined full-width-half-maximum (FWHM) from the MgII line for each quasar. The BH mass uncertainties from the single-epoch BH mass estimator with the MgII line is roughly 0.4\, dex \citep{shen_etal_2023}, though it might be larger for the CUBS quasars since they are outliers in luminosity compared to the quasar population used in reverberation mapping studies. The virial BH masses and the luminosity at 3000\,\AA\ (as functions of redshift) are plotted in Figure \ref{fig:quasar} and tabulated in Table \ref{tab:quasar}. 

\subsection{Radio Properties}
We obtain the radio properties of the quasars by cross-matching with the Rapid Australian Square Kilometre Array Pathfinder (ASKAP) Continuum Survey (RACS) DR1 catalog \citep{Hale_etal_2021_RACS}. 
ASKAP surveys the sky south of $+41\deg$ declination at a central frequency of 887.5\,MHz with a spatial resolution of $\sim$15$''$ (convolved to 25$''$ for the source catalog). 
We first match the quasar positions in the RACS DR1 catalog for all sources within 30$''$. 
A total of seven quasars have radio counterparts; two (J0110, J2245) show extended/lobed detection (lobe-dominated), and the rest (J0114, J0248, J0357, J0454, J2339) show unresolved single detection (core-dominated or unresolved). 
For the lobed-dominated sources, we use the total source flux as the radio flux. We calculate the radio-loudness ($R$) as the ratio of flux density at 6\, cm and 2500\,\AA\, in the rest frame, $R=f_{\rm \nu, 6 cm}/f_{\rm \nu, 2500\,\AA}$ \citep{Stocke_etal_1992}. 
The UV flux density $f_{\rm \nu, 2500 \AA}$ is calculated from the best-fit quasar spectra from PyQSOFit. 
The rest-frame 6\,cm flux density is extrapolated from the RACS radio flux by assuming a spectral index $\alpha$ ($F_{\rm \nu}\propto\nu^{\alpha}$) of $-0.5$ . 
The RACS DR1 catalog requires a 5$\sigma$ detection for sources to be included in the catalog and has an overall 95$\%$ point source completeness at an integrated flux density $\approx3$\,mJy. 
Adopting 3\,mJy as the upper limit in radio flux density, we find most of the non-detected sources have an upper limit of $R\lesssim10$. Typically, quasars with $R\gtrsim10$ are defined as radio-loud quasars \citep{Kellermann_etal_1989}. We identify five radio-loud quasars, and the remaining quasars are undetected or radio-quiet.
For the quasars above declination of $-40$\, degrees, we also cross-check with the FIRST and VLASS surveys and find consistent radio properties at 3\,cm. 
The radio-loudness parameter $R$ is tabulated in Table \ref{tab:quasar}.

\begin{figure*}
\raggedright
    \includegraphics[width=0.24\textwidth]{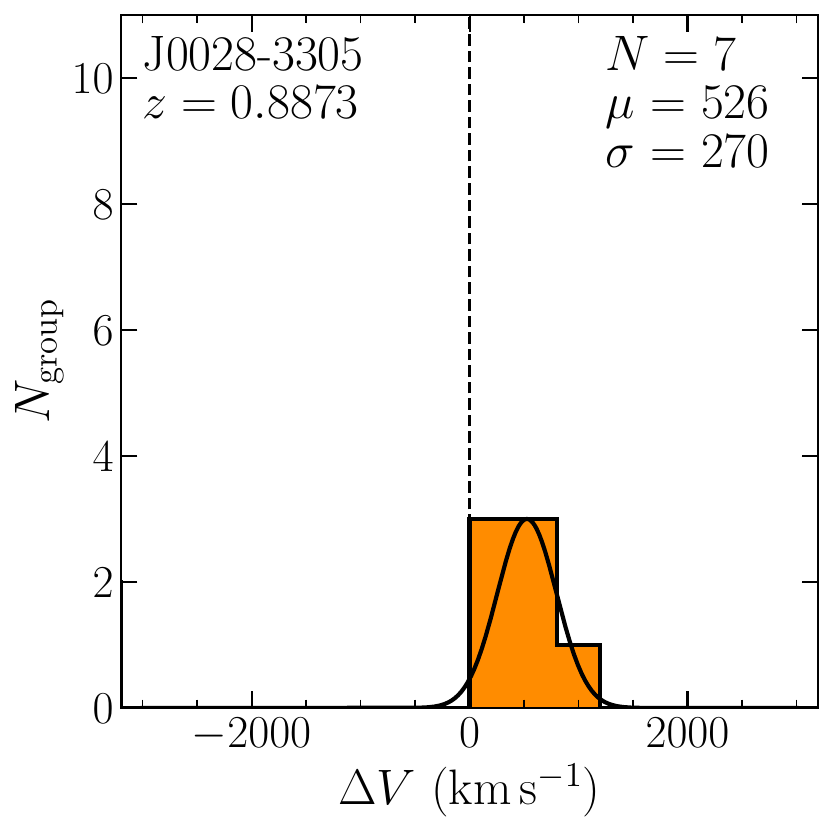}
    \includegraphics[width=0.24\textwidth]{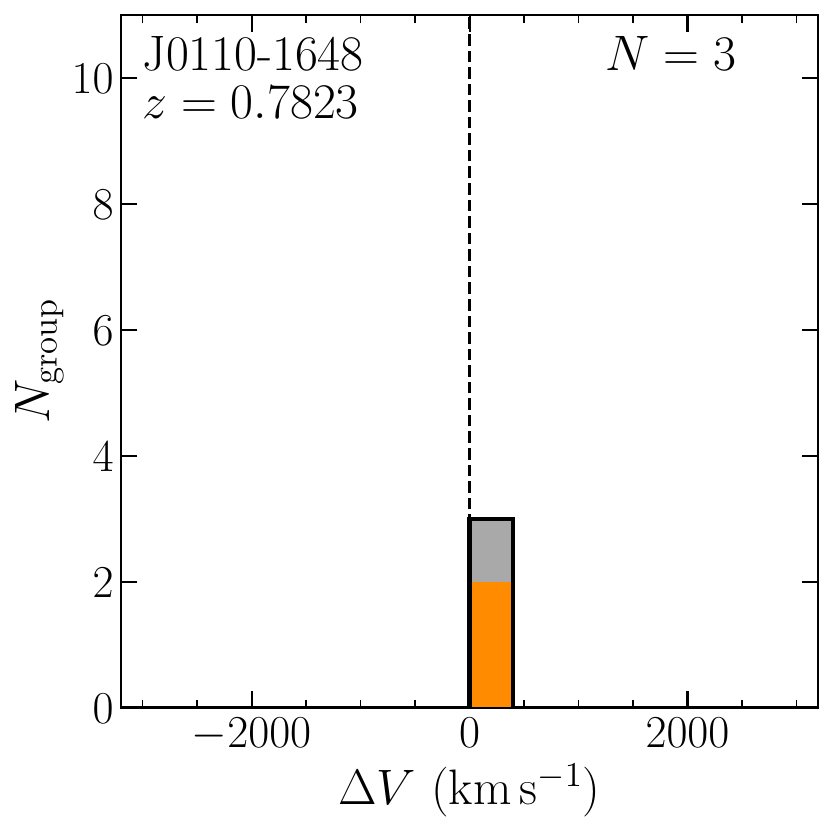}
    \includegraphics[width=0.24\textwidth]{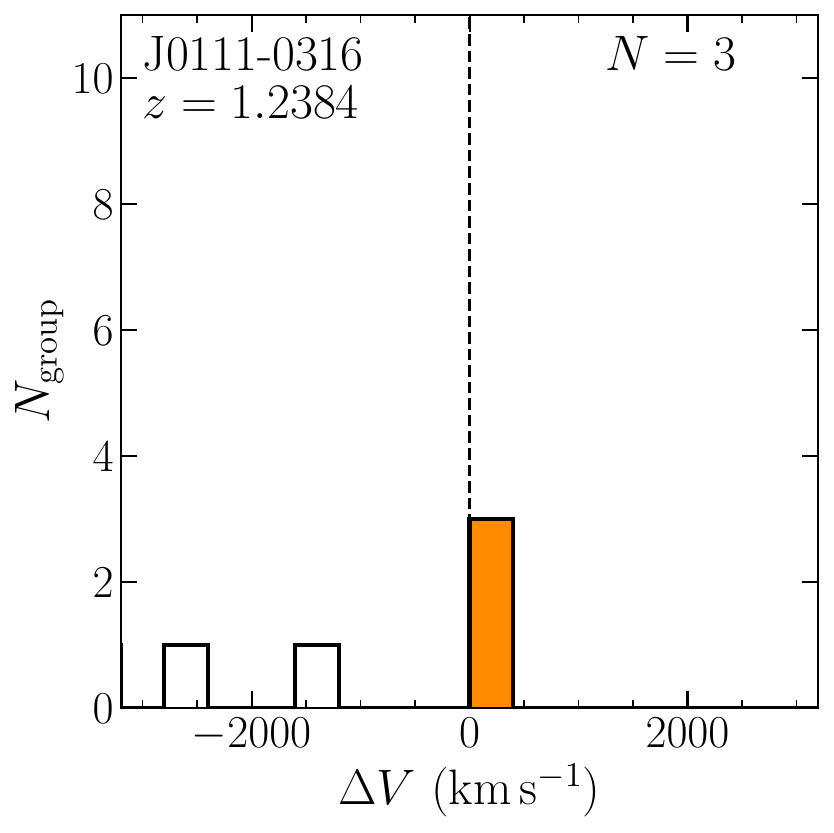} 
    \includegraphics[width=0.24\textwidth]{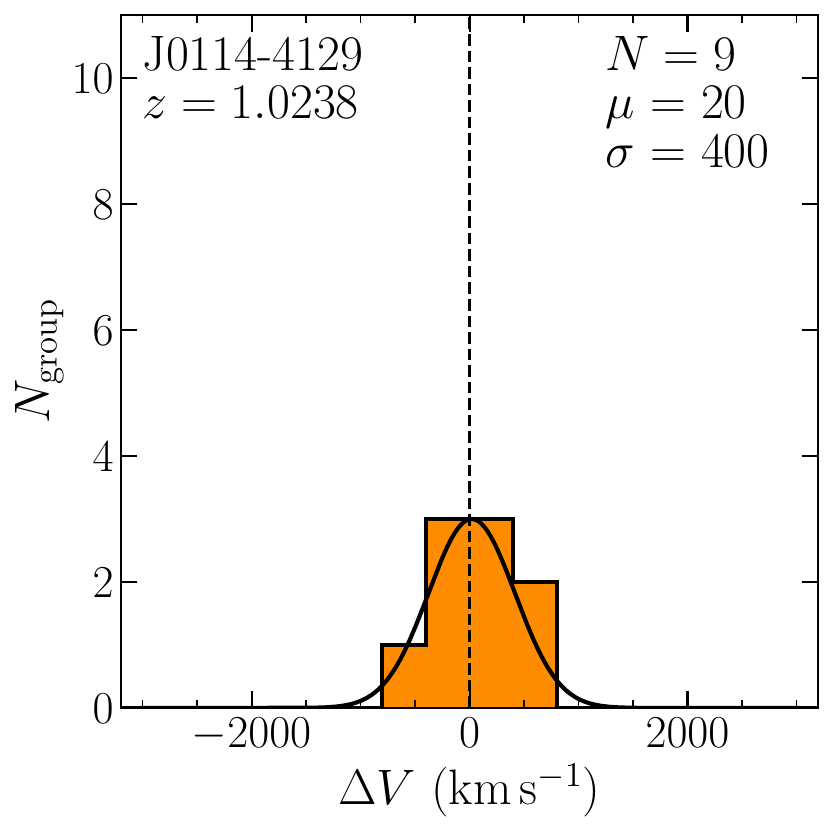} \\
    \includegraphics[width=0.24\textwidth]{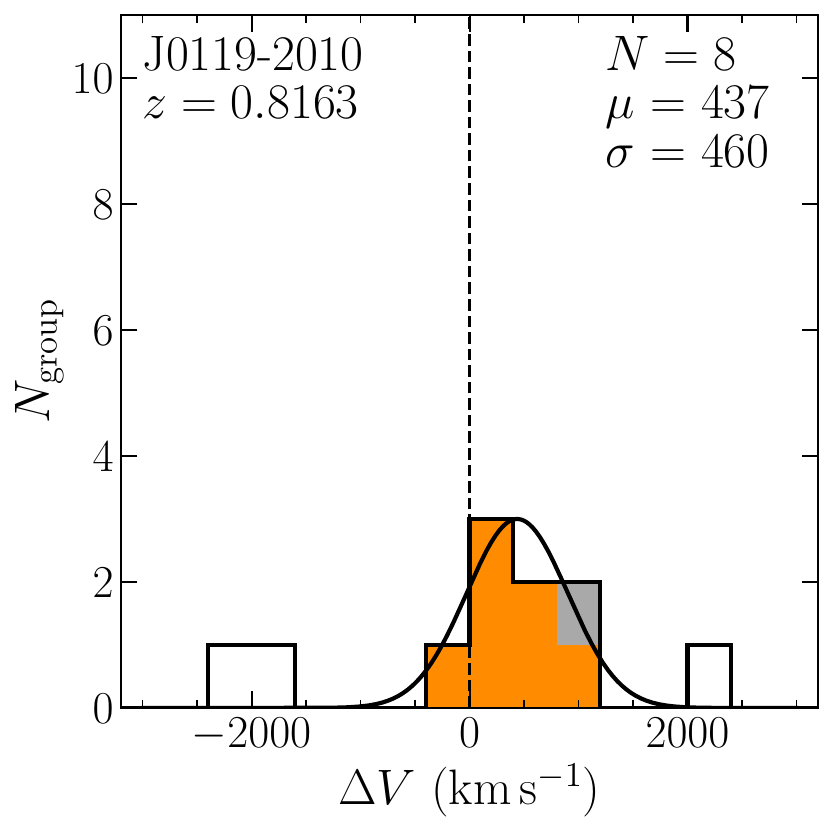}
    \includegraphics[width=0.24\textwidth]{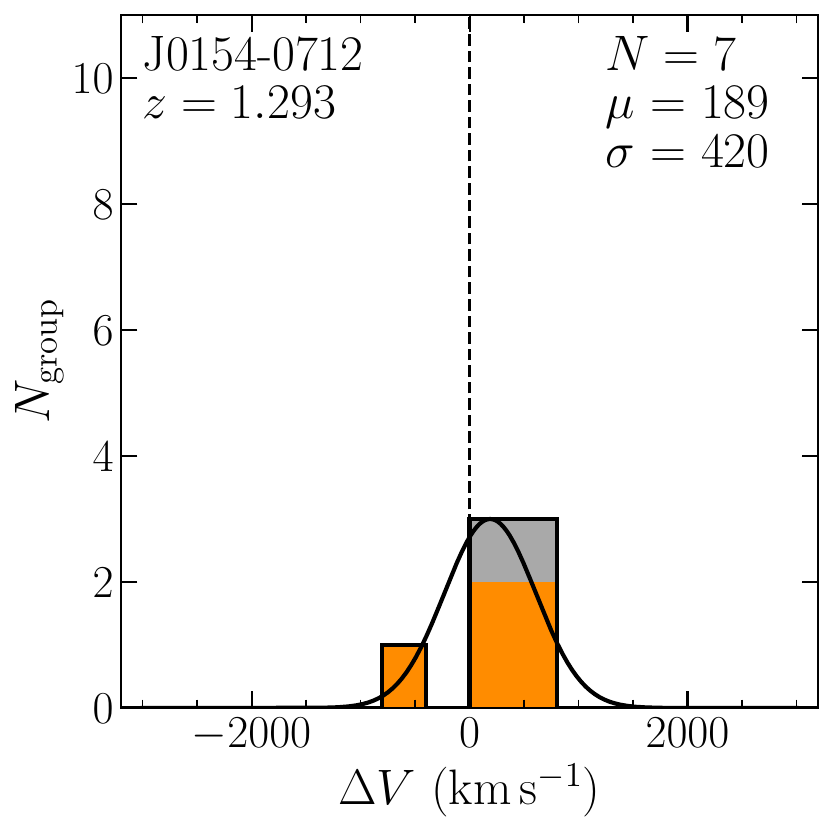} 
    \includegraphics[width=0.24\textwidth]{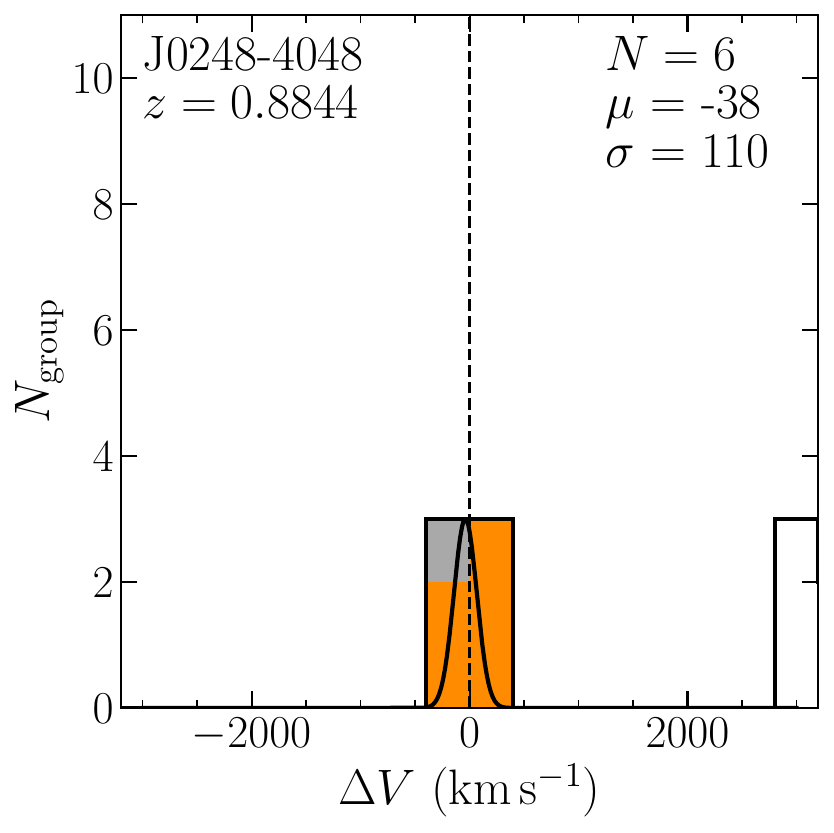}
    \includegraphics[width=0.24\textwidth]{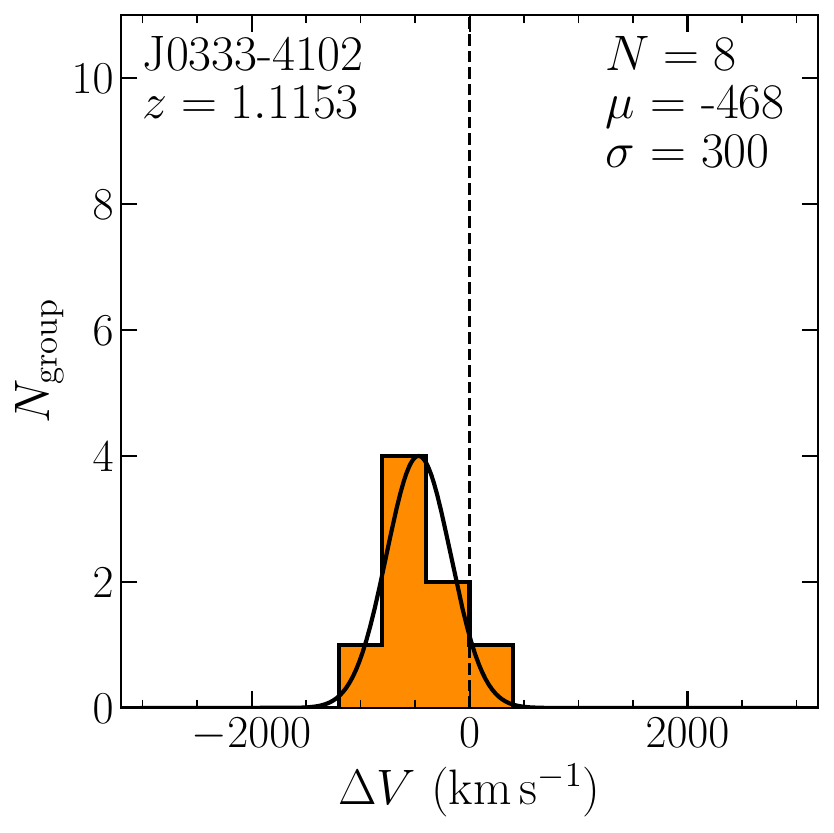} \\
    \includegraphics[width=0.24\textwidth]{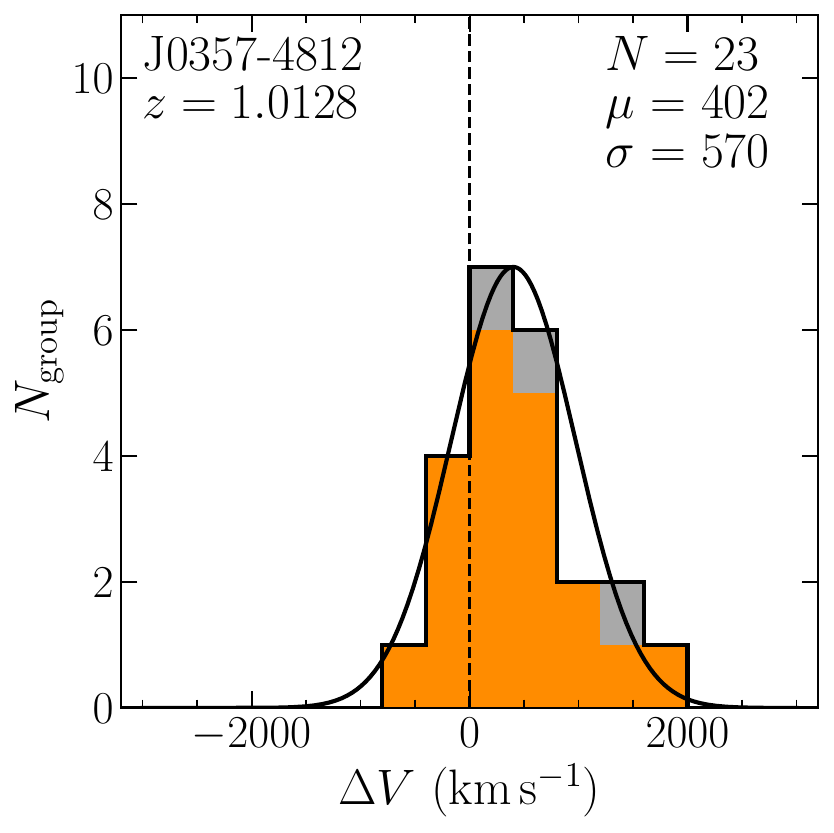} 
    \includegraphics[width=0.24\textwidth]{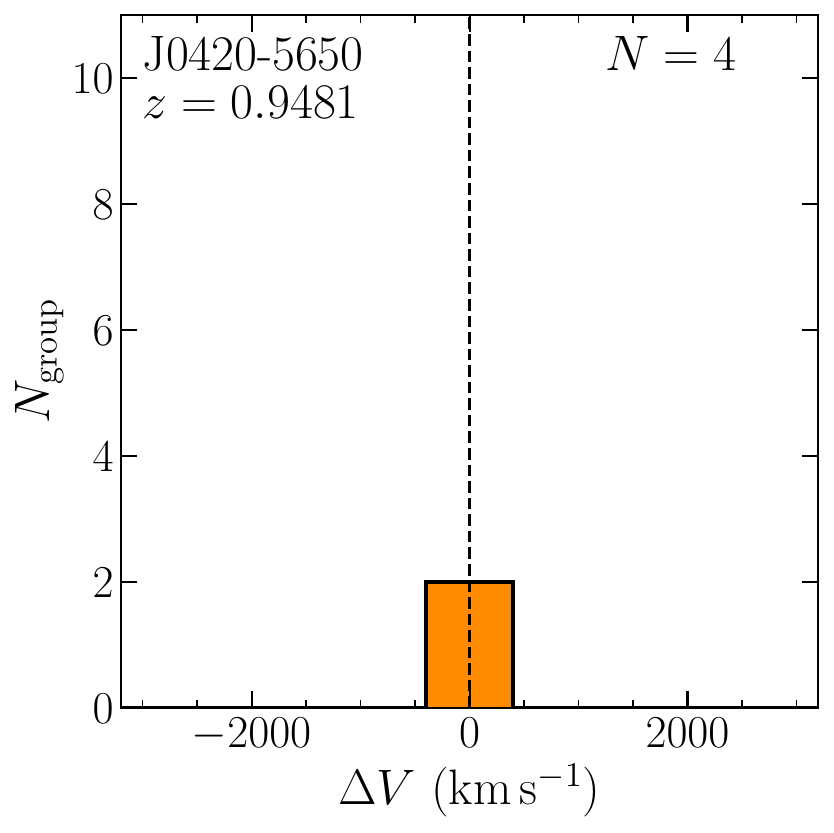}
    \includegraphics[width=0.24\textwidth]{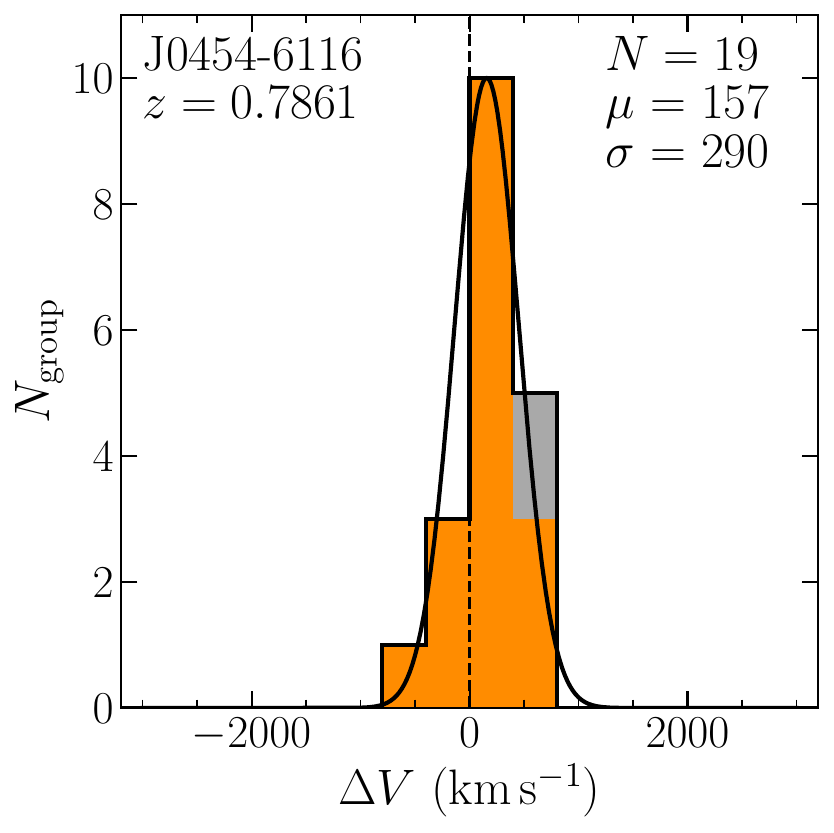}
    \includegraphics[width=0.24\textwidth]{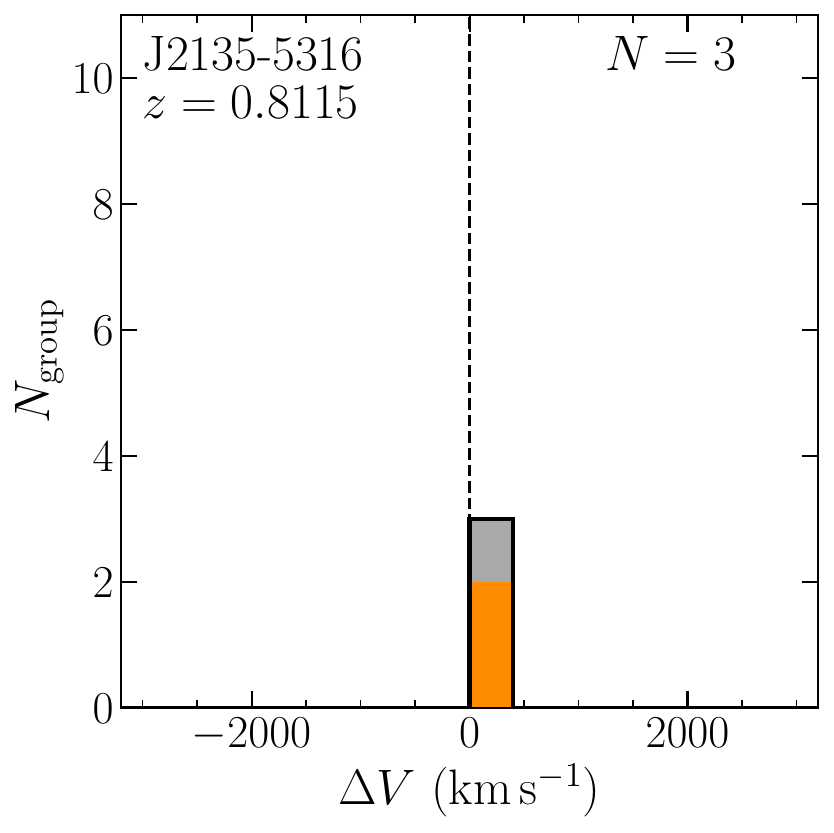} \\
    \includegraphics[width=0.24\textwidth]{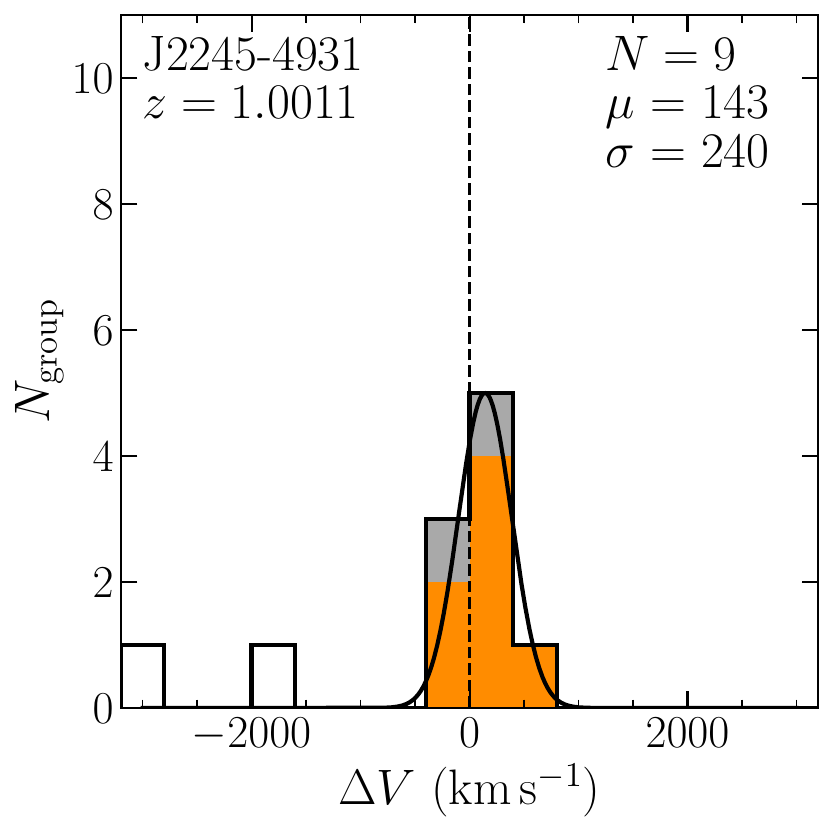}
    \includegraphics[width=0.24\textwidth]{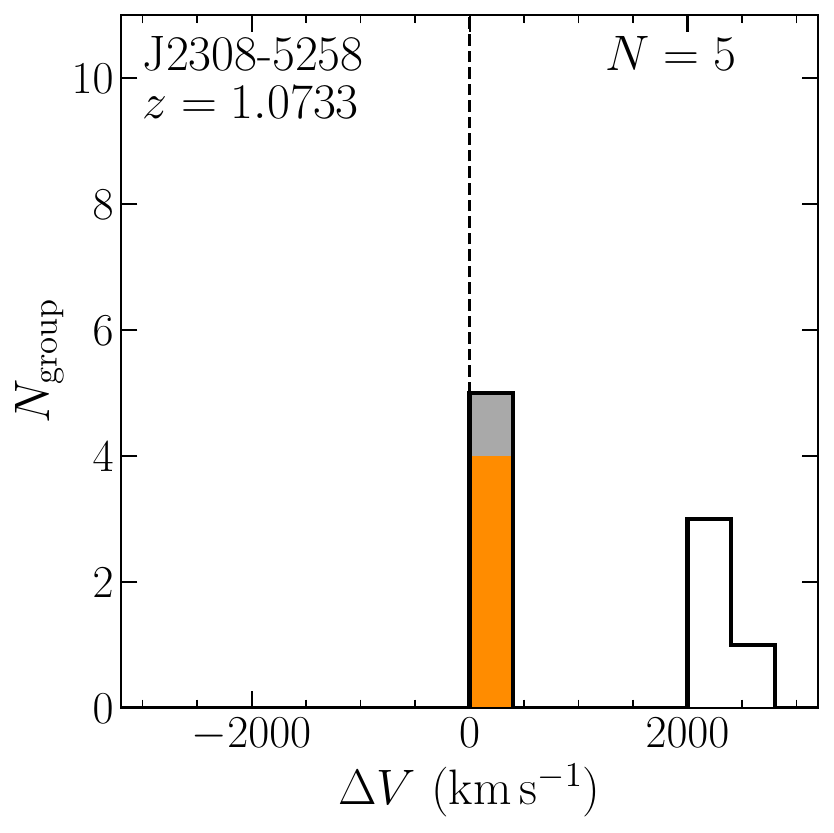}
    \includegraphics[width=0.24\textwidth]{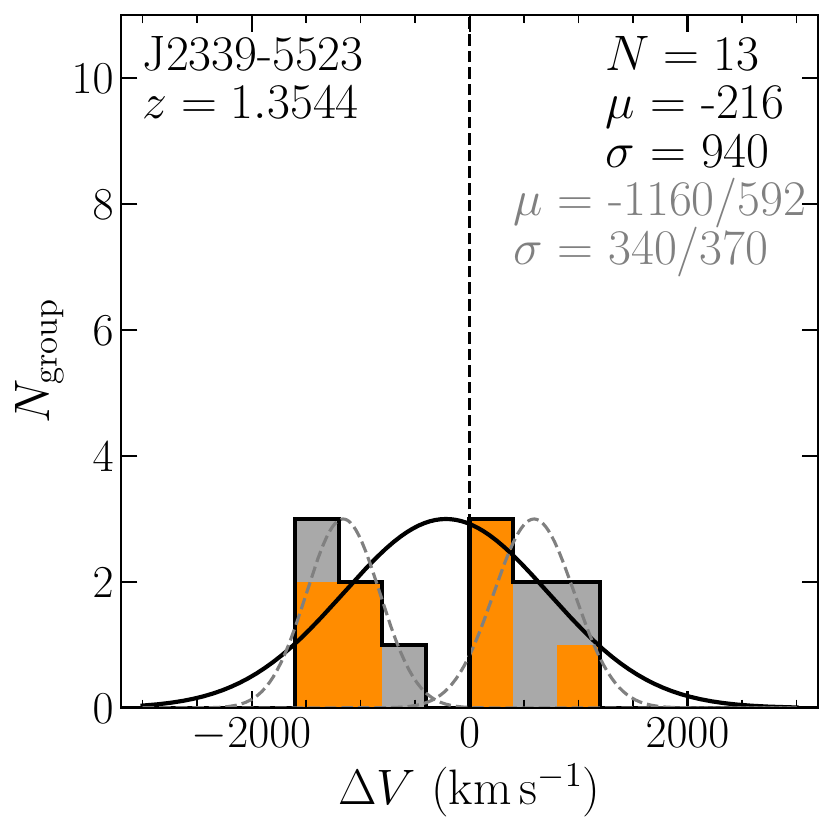} \\
\caption{Velocity distribution of galaxies within each quasar field, with the quasar at $\Delta\,V=0$. The black histogram shows all galaxies around the quasar redshift, and the orange (grey) histograms show the double(single)-line redshift within each group identified by the selection criteria described in Section \ref{sec:group}. For galaxy groups with more than 5 galaxies, we fit the velocity distribution with a normal distribution to measure the mean velocity $\mu_{{\rm Vel}}$ and velocity dispersion $\sigma_{{\rm Vel}}$. The best-fit normal distributions (black solid lines) are scaled to the peak of the velocity distribution. The bimodal fit for the field J2339 is displayed in grey dashed lines.}
\label{fig:vel_hist}
\end{figure*}

\section{Results}\label{sec:results}

\subsection{Group Environment around CUBS quasars}\label{sec:group}

To identify member galaxies in the quasar galaxy groups, we search for galaxies close to the quasar redshift in the MUSE field of view. We calculate the group mean velocity ($\mu_{{\rm Vel}}$) and velocity dispersion ($\sigma_{{\rm Vel}}$) by fitting the galaxy relative velocity within $|\Delta V|<1500$\,km\,s$^{-1}$ of the quasar redshift with a normal distribution. The initial velocity range ($|\Delta V|<1500$\,km\,s$^{-1}$) roughly equals the velocity dispersion of the most massive galaxy clusters. For more robust identification, we re-calculate the new $\mu_{{\rm Vel}}$ and $\sigma_{{\rm Vel}}$ after sigma-clipping at 3$\sigma$ and remove all sources beyond 3$\sigma$ of the original $\mu_{{\rm Vel}}$ and $\sigma_{{\rm Vel}}$, this calculation is repeated for 2-3 times until the group member identification has converged. Figure \ref{fig:field} shows the MUSE white-light images and the galaxies identified within each quasar group. Figure \ref{fig:vel_hist} shows the velocity distribution of each group. All galaxies within $|\Delta V|<3000$\, km\,s$^{-1}$ of the quasar are shown in open histograms, and the identified group members shown in orange (double-feature redshifts) and grey (single-feature redshifts) histograms.

The groups include 3--23 galaxies around the quasars (including the quasar host galaxy), indicating the CUBS quasars reside in poor to massive groups. For groups with $N_{\rm gal}>5$, we report the final $\mu_{{\rm Vel}}$ and $\sigma_{{\rm Vel}}$ of the group. The velocity dispersion ranges from 150--600\, km\,s$^{-1}$ (except for J2339), which suggests our initial velocity search range is sufficient. While there could still be chance projections in the identified group members, the majority of the galaxies in $N_{\rm gal}>5$ groups are projected within the virial radii, estimated with the derived halo masses (see Section \ref{sec:bh-gal-halo}). In addition, most of the identified group members are likely bound in $N_{\rm gal}>5$ groups, their relative velocities to the quasars are less than the escape velocities at the corresponding galaxy distances. Removing the few galaxies near the virial radius or above to the escape velocity does not change the estimated velocity dispersion or halo mass significantly. The final group size, mean velocity, and velocity dispersion are tabulated in Table \ref{tab:group}. 

The velocities of galaxies in the field of J2339 show a bimodal distribution, suggesting there might be two individual galaxy groups perhaps starting or undergoing a merger event. When fitted with two normal distributions, we find two distinct groups of 7 (closer to the quasar redshift) and 6 galaxies with similar velocity dispersion of $\approx$ 350\, km\,s$^{-1}$.  While $\sim$1000\, km\,s$^{-1}$ {($\sigma_{\rm Vel}$ of J2339 fitted with a single Gaussian)} is still within a reasonable range for a galaxy group/cluster, $\approx$300\, km\,s$^{-1}$ is more consistent with other groups and better describes the bimodal velocity distribution. We report the group properties of the group closer to the quasar redshift in Table \ref{tab:group} and plot both velocities in the remaining figures of this paper. 

\begin{figure}
\centering
    \includegraphics[width=0.45\textwidth]{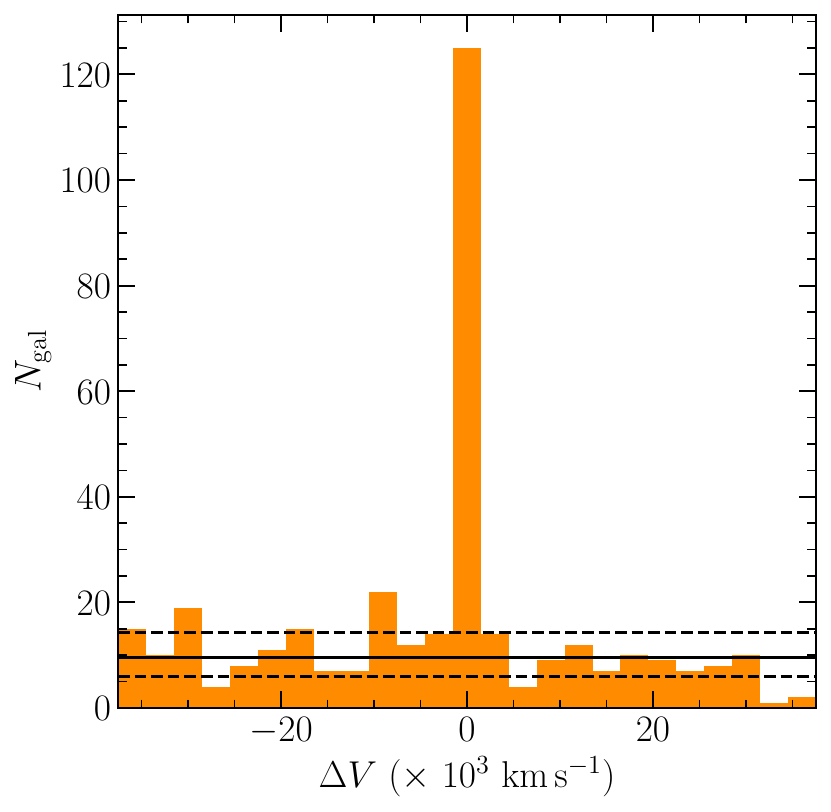}
\caption{Stacked distribution of the relative velocity from the central quasars. The solid (and dashed) line shows the median (16th and 84th percentiles) number count of the background galaxies (i.e., excluding the central bin).}
\label{fig:stack_dvel}
\end{figure}

\subsection{Overdensities} \label{sec:overdensity}
To study the quasar environments, we search for overdensities around the central quasars using the relative velocity distribution. Following \cite{Stott_etal_2020}, we calculate the overdensity $\delta_{g}$ using the equation: 
\begin{equation}
    \delta_{g} = \frac{N_{\rm group}-N_{\rm bkg}}{N_{\rm bkg}}
\end{equation}
where $N_{\rm group}$ is the number of objects in each quasar group and $N_{\rm bkg}$ is the expected number of background objects without the presence of any structure. We calculate $N_{\rm bkg}$ by stacking the relative velocity distribution for all quasar fields with a bin size of $3000$\,km\,s$^{-1}$ (i.e., $\pm1500$\,km\,s$^{-1}$), centered on the quasar systematic velocity. The velocity {bin size} is chosen to ensure all identified group galaxies are within the center bin of the stacked histogram (see Figure \ref{fig:vel_hist}). The {full $\Delta {V}$ range} of $\pm\,36,000$\,km\,s$^{-1}$ shown in Figure \ref{fig:stack_dvel} roughly corresponds to 0.2--0.3 in $\Delta z$ depending on the quasar redshift.

Figure \ref{fig:stack_dvel} shows the stacked distribution of relative velocity from the central quasars. The median (16th, 84th percentiles) number counts of the background galaxies are $N_{\rm bkg}=9.5$ (6.0, 14.3) when excluding the central bin. The stacked $\delta_{g}$ across all 15 quasar fields is 12.4$^{+1.2}_{-1.1}$ (i.e., $\approx11\sigma$), and the individual $\delta_{g}$ of each quasar field ranges $\approx$4--35 (tabulated in Table \ref{tab:group}) using the galaxy counts in each field and the {\it average} background galaxy count across 15 fields, $N_{\rm bkg}=0.63$. The uncertainties of $\delta_{g}$ are calculated through the Bayesian approach described in \cite{Kraft_etal_1991} for determining the Poisson confidence level for a number count given by a background. Following the convention in the literature, we find 11 quasars reside in $>$2$\sigma$ overdense environments (i.e., galaxy groups with $\gtrsim 5$ members), and 5 quasars reside in $>$3$\sigma$ overdensity ($\gtrsim 9$ members). When excluding the ``single-feature'' redshifts, the results remain unchanged; the stacked $\delta_{g}$ is 14.9$^{+1.4}_{-1.3}$ (i.e., $\approx11\sigma$), and the same 11(5) quasars are above the $>2\sigma(3\sigma)$ threshold. While $N_{\rm bkg}$ and $\delta_{g}$ depend on the exact choice of the velocity grid, the significance of overdensity ($\sigma_{\delta}$) remains consistent with bin sizes of $2500-4500$\,km\,s$^{-1}$. We note that the decline at $\Delta {V}>25,000$\, km\,s$^{-1}$ in the stacked relative velocity (Figure \ref{fig:stack_dvel}) could be due to the lower completeness in redshift measurement at $z>1$. {However, the overdensity measurements are not sensitive to the exact choice of the full $\Delta {V}$ range, provided that it is large enough for accurately estimating the background galaxy counts. For our sample, the median background galaxy count remains consistent at $N_{\rm bkg}=9-10$ with full $\Delta {V}$ ranges of $\pm\,24,000-36,000$\,km\,s$^{-1}$.}

The MUSE FOV (1$'\approx500$\, kpc at $z\approx1$) might not be large enough to survey all galaxies in the galaxy groups/clusters. We find 1$-$3 additional galaxy group members within a projected distance of 500\, kpc around half of the CUBS quasars in the LDSS3/IMACS redshift surveys. These additional galaxies are all found in the fields of the $>2\sigma$ overdense quasars and the measured velocity dispersion is consistent within 10\% if these galaxies are included. The only two exceptions are the fields J0114 and J0119. The additional galaxy found in the J0114 field lies at the edge of the velocity distribution and thus increased the velocity dispersion by $\approx$30\% when included. For the field of J0119, six additional galaxies were found in the LDSS3 FOV near the quasar redshift, which made J0119 one of the most overdense groups in the sample. The estimated velocity dispersion of the galaxy group around J0119 decreases by $\approx$20\% when including the six additional galaxies. However, the source selection of the LDSS3/IMACS redshift surveys is based on color selection to prioritize galaxies in the quasar foreground and the redshift surveys are much shallower than the MUSE observation. Therefore, it is non-trivial to quantify the selection effect in overdensity estimation, and we choose to focus on the overdensity in the MUSE FOV only.

\subsection{Stellar Masses of the Group Members}

We estimate the stellar masses for the galaxies within the CUBS quasar groups by galaxy spectral modeling with the code {\tt BAGPIPES} \citep{Bagpipes1,Bagpipes2} using both spectroscopic and photometric observations. We extract the galaxy spectra within a $\approx$0.6\,$''$\,radius using {\tt mpdaf}, i.e., the same spectra as used for the redshift measurements. We extract deep psuedo-photometry in $g$, $r$, and $i$-band by convolving the MUSE datacubes with the $g$, $r$, $i$ psuedo-bandpasses {(i.e., boxcar functions around 4800--5800, 6000--7500, and 7500--9000\,\AA, following previous CUBS papers).}
For bright galaxies ($i\gtrsim$\,20.5\,mag), we supplement with the $g$, $r$, $i$, $z$, $Y$-band photometry from the Dark Energy Survey Y3 GOLD catalog \citep{DES_Y3}. Finally, we include the $H$-band photometry from our Four Star observations. For the MUSE and $H$-band photometry, we first perform {\tt Source Extractor} on the pseudo-$r$ band image and use the pseudo-$r$ band segmentation map to extract photometry from each MUSE and $H$-band images to ensure the photometry is obtained from the same region. The photometric uncertainties are set to be 10\%.

{\tt BAGPIPES} is a Bayesian fitting code that can model galaxy spectra with spectroscopic and photometric data simultaneously. We model the galaxy spectra with the \cite{Kroupa_2001} initial mass function, \cite{BC03} stellar population models, an exponential star formation history, and include nebular emission and dust extinction models from \cite{Calzetti_etal_2000}. To fit spectroscopic and photometric data simultaneously, we allow zeroth to second order calibrations and a noise scaling parameter that accounts for relative flux calibration as suggested by {\tt Bagpipes}, e.g., corrections for aperture or template mismatch, underestimation of uncertainties, and wavelength-dependent flux calibration in the spectroscopic data. As a {robust check}, we tested different star formation history (exponential and double power law) and dust extinction models (\cite{Calzetti_etal_2000} and \cite{Charlot_Fall_2000}) and found the estimated galaxy masses to be consistent regardless of the model and parameter setup. 

Galaxies closest to the central quasar are often heavily contaminated, or even out-shined, by the quasar light. For galaxies very close to the quasars, we perform quasar light subtraction using the methods from \cite{Johnson_etal_2018} and \cite{Helton_etal_2021} to recover the quasar-subtracted spectra and photometry for the galaxies closest to the quasars. In short, we model the light around the quasars using a linear combination of galaxy and quasar eigenspectra to decompose the quasar and galaxy light. For the $H$-band image, we use {\tt GALFIT} \citep{GALFIT} to perform 2D surface brightness decomposition to remove the central quasar light. Finally, we follow the same procedures described previously to extract MUSE spectra, pseudo-photometry, and $H$-band photometry from the quasar-subtracted datacubes and images. Of the 11 galaxies that are strongly contaminated by the quasar light, we were able to recover the galaxy photometry and stellar mass for four galaxies. 

The derived properties of the galaxies in each quasar group are tabulated in Table \ref{tab:galaxy}, along with their relative velocities and projected distances from the central quasar. The estimated galaxy masses are $8.5<{\rm log({M}_{\rm \star}/{M}_{\odot})}<11.6$ , with median ${\rm log({M}_{\rm \star}/{M}_{\odot})}=9.9$, and the typical mass uncertainty from {\tt Bagpipes} is $\approx$0.1\,dex, excluding systematic uncertainties. 

\section{Discussion}\label{sec:discussion}

\subsection{BH--galaxy--halo relation}\label{sec:bh-gal-halo}

In this section, we discuss the relations between the central SMBH, the host galaxy, and the surrounding galaxy cluster/group. Observations of local and intermediate ($z<2$) redshift quasars have revealed tight correlations between the SMBH mass and host galaxy mass \citep[e.g.,][]{Jahnke_etal_2009, Bennert_etal_2011}, luminosity \citep[e.g.,][]{Laor_1998, Peng_etal_2006a, Decarli_etal_2010}, velocity dispersion \citep[e.g.,][]{Treu_etal_2004, Woo_etal_2006, Woo_etal_2010}, as well as halo mass ($M_{\rm h}$) and temperature \citep[e.g.,][]{Ferrarese_2002, Gaspari_etal_2019, Donahue_Voit_2022}.
The correlation between BH, host galaxy, and dark matter halo can be explained through self-regulated feedback processes, i.e., galaxy growth halts as the cool, star-forming ISM gas heats up or gets expelled to CGM scale when the central SMBH releases energy into the surrounding galaxy and halo. If the BH mass is controlled by the halo binding energy, then $M_{\rm BH} \propto M_{\rm h}^{5/3}$ is predicted \citep{Silk_Rees_1998, Booth_Schaye_2010}. Though some works argue against a direct correlation between $M_{\rm BH}$ and $M_{\rm h}$ \cite[e.g.,][]{Kormendy_Ho_2013}. In addition, through abundance matching and direct measurements of galaxy clusters, we can probe the efficiency of turning baryonic matter into stars using the ratio of galaxy stellar and halo masses. Maximum star formation efficiency occurs around halo masses of typical $L_{*}$ galaxies {(${M}_{\rm h}\sim10^{12}{M}_{\odot}$)}, and star formation efficiency decreases at both low and high mass ends due to strong feedback from star formation and AGN activities \citep[e.g.,][]{Behroozi_etal_2013, Kravtsov_etal_2018, Wechsler_Tinker_2018}.

\begin{figure}
\centering
    \includegraphics[width=0.45\textwidth]{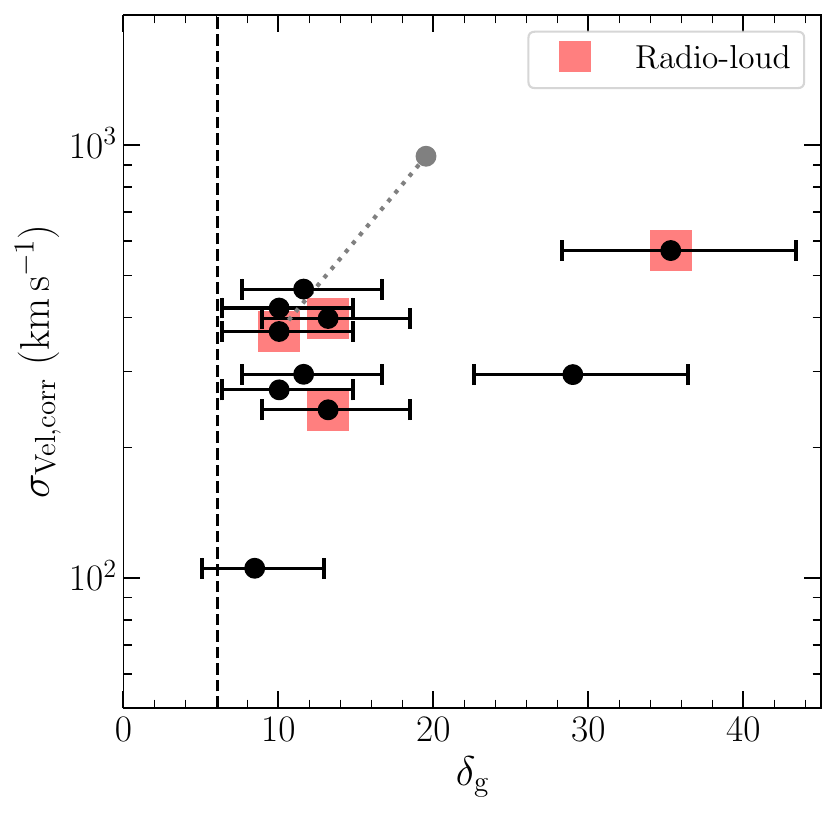}
\caption{Velocity dispersion of the galaxy groups as a function of overdensity. The velocity dispersion is only calculated for galaxy groups with more than five galaxies. The radio-loud quasars are labeled with red squares around the data points. The grey point shows the overdensity and velocity dispersion of the galaxy group around J2339 when the velocity distribution is fitted with a single Gaussian distribution, and is connected to the two Gaussian measurements with a dotted line. The vertical dashed line shows the $2\sigma$ overdensity limit. The velocity dispersion is expected to correlate positively with overdensity, as velocity dispersion traces the dynamical mass of the dark matter halo around the central quasar. However, the correlation is not statistically significant for our sample.}
\label{fig:od_vdisp}
\end{figure}

\subsubsection{Galaxy and halo mass estimation}

We estimate the {bulge (or galaxy)} stellar masses of the quasar hosts to be around $10^{11}$--$10^{12.5} {\rm M}_{\odot}$ using the $M_{\rm BH}-M_{\rm \star, bulge}$ relation from \cite{Kormendy_Ho_2013}. For the purpose of this work, the difference between host galaxy mass and bulge mass is negligible. For all 15 quasar fields, the quasar host galaxies are more massive than other group members and encompass $\approx 50\%-99\%$ of the total stellar mass. Therefore, the quasar host galaxies are very likely the central galaxies in each of the galaxy groups. Central galaxies of galaxy groups, similar to brightest cluster galaxies (BCGs) in galaxy clusters, hold a special position in structure formation as they are centered in the group/cluster halo \citep[e.g.,][]{Lin_Mohr_2014}.

\begin{figure*}
\centering
    \includegraphics[width=0.45\textwidth]{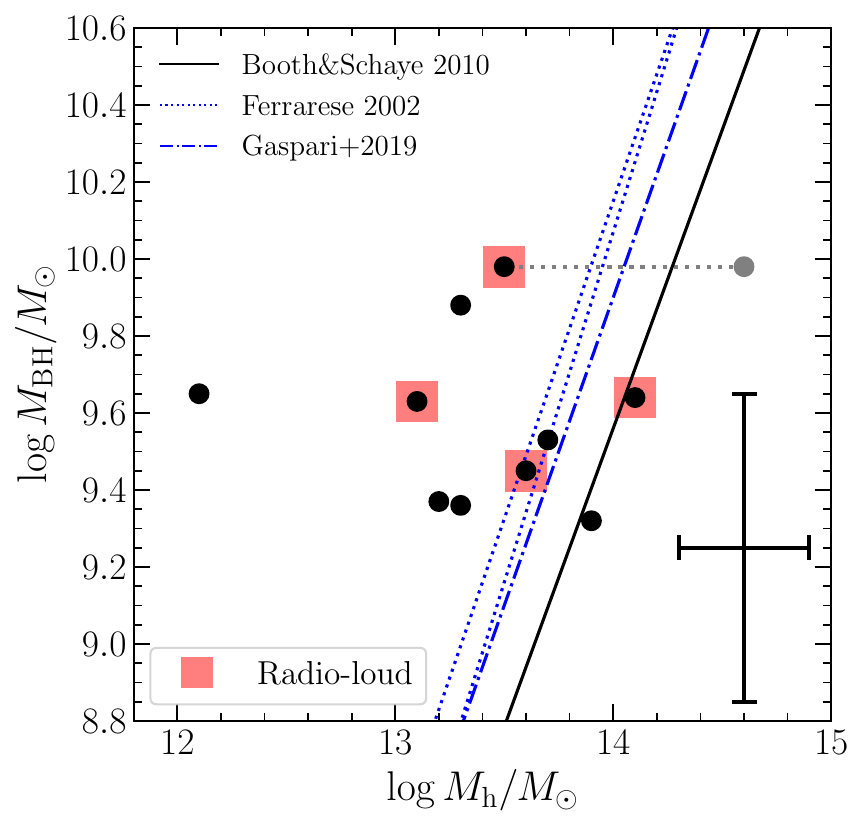}
    \includegraphics[width=0.45\textwidth]{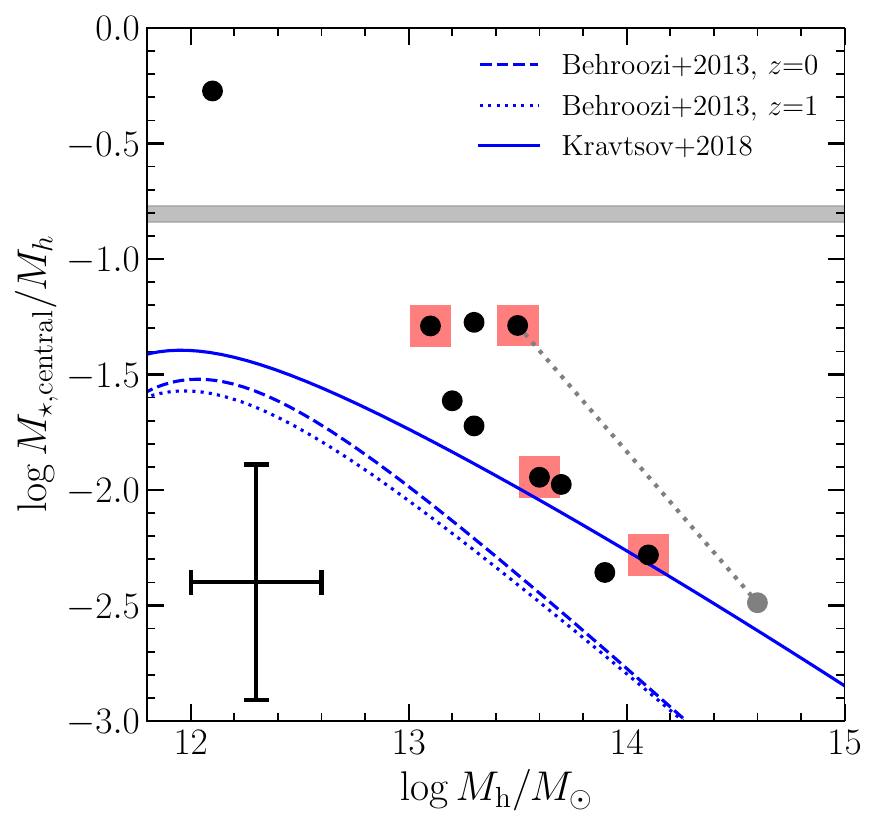}
\caption{BH mass (left) and stellar mass-to-halo mass ratio (right) as functions of halo mass. 
In the left panel, the lines show the best-fit relations from local observations (blue dotted lines: Equation 4 and 6 from \citep{Ferrarese_2002}, blue dotted-dash line: \cite{Gaspari_etal_2019}) and the theoretical prediction from \cite{Booth_Schaye_2010} (black solid line). In the right panel, the blue solid line shows the \cite{Kravtsov_etal_2018} stellar mass-to-halo mass relation from direct observations and abundance matching at $z<0.1$, and the blue dashed and dotted lines show the relations at $z\approx0$ and $z\approx1$ from \cite{Behroozi_etal_2013}. {J0248, which lies above the average stellar mass-halo mass relations and the observed cosmic baryon fraction \citep[grey shaded area, ][]{Planck_2018}, could be an outlier in the BH-galaxy-halo relations or have less accurate halo mass estimation (see Section 5.1.2).} The typical uncertainties (dominated by systematic uncertainties) are shown in each panel. There is little redshift evolution in the stellar mass-to-halo mass relation. Most of our sample lies above the average stellar mass-halo mass relations from abundance matching and theoretical prediction, hinting that the luminous CUBS quasars may represent outliers in the BH-galaxy-halo relations or have less reliable virial BH masses.}
\label{fig:mhalo_ms_mhalo}
\end{figure*}

Figure \ref{fig:od_vdisp} shows the relation between the velocity dispersion and the overdensity parameter. The velocity dispersion traces the dynamical mass of the dark matter halo around the central quasar and the surrounding neighbors, therefore the velocity dispersion is expected to  correlate positively with {the overdensity and the number of group galaxies}. Though the correlation is not statistically significant in our sample  (the Pearson correlation coefficient is 0.39, with a p-value of 0.26, for the groups with $N_{\rm gal}>5$).
Assuming the galaxy groups are virialized, the halo mass can also be traced dynamically from the velocity dispersion in the galaxy groups. Following Equation 1 from \cite{Munari_etal_2013}, we convert the line-of-sight velocity dispersion to halo mass:
\begin{equation}\label{eq1}
    \frac{{\rm \sigma}_{\rm Vel}}{\rm km\,s^{-1}} = A \left[ \frac{h(z){\rm M}_{\rm h}}{10^{15}{\rm M}_{\odot}} \right] ^\alpha, 
\end{equation}
where $A = 1177$ and $\alpha=0.364$ for their simulations using galaxies as tracers and including AGN feedback. The halo mass here refers to $M_{\rm 200}$, the dark matter halo mass within $R_{\rm 200}$, where the average density is 200 times the critical density of the Universe.

In the small $N_{\rm gal}$ regime, all velocity dispersion estimators are statistically biased by small number statistics. We calculate the halo mass for groups with more than 5 members using ${\rm \sigma}_{\rm Vel}$ and follow the parametric corrections for small $N_{\rm gal}$ from \cite{Ferragamo_etal_2020} to derive the unbiased standard deviation ${\rm \sigma}_{\rm Vel, corr}$. {The intrinsic scatter of Equation \ref{eq1} is relatively small ($\approx5\%$), so the halo mass uncertainty ($\approx0.1-0.3$\, dex) is dominated by the measurement uncertainty of ${\rm \sigma}_{\rm Vel}$ (assuming $\approx10-30\%$).} {We also note that there are additional biases in the halo mass estimation from incomplete sampling of member galaxies and interloper contamination \citep{Ferragamo_etal_2020}, which are relatively small and thus not included in our calculation.}
The estimated halo mass ranges from $10^{13}-10^{14} {\rm M}_{\odot}$, suggesting the most luminous quasars reside in more massive dark matter halos than the general quasar population. In contrast, observations show typical optically-selected quasars reside in $M_{\rm h} \sim 10^{12}-10^{13} {\rm M}_{\odot}$ \citep[e.g.,][]{Hickox_etal_2009}. Results from halo occupation modeling and clustering analyses also showed that the median masses of quasar halos are $M_{\rm h}=4.1\times10^{12}\,{\rm M}_{\odot}$ for central quasars at $z\approx1.4$, and that halos with $M_{\rm h}\sim10^{13}\,{\rm M}_{\odot}$ are $\approx1\sigma$ deviation from the full mass distribution of central quasar halos \citep{Richardson_etal_2012}.

\subsubsection{$M_{\rm BH}-M_{\rm h}$ and $M_{\rm \star}/M_{\rm h}-M_{\rm h}$ relations}

Figure \ref{fig:mhalo_ms_mhalo} (left panel) shows the relation between the halo mass and the BH mass, along with the theoretical prediction of $M_{\rm BH} \propto M_{\rm h}^{1.55}$ from \cite{Booth_Schaye_2010} and best-fit relations from \cite{Ferrarese_2002} and \cite{Gaspari_etal_2019}. For the \cite{Gaspari_etal_2019} relation, we follow \cite{Voit_etal_2023} and convert $M_{\rm 500}$ derived from the halo gas temperature to $M_{\rm h}$ assuming a Navarro-Frenk-White mass profile with a concentration parameter $c_{200}\approx4$. Despite the limited dynamical range in BH mass and large scatter, the BH mass and halo mass for the majority of the sample are roughly consistent with the local observed relations and the predicted $M_{\rm BH} \propto M_{\rm h}^{5/3}$ relation.

The J0248 galaxy group appears to be an outlier of the sample, the velocity dispersion is much smaller than other galaxy groups with a similar number of galaxies, and the halo mass $\sim10^{12.1} {\rm M}_{\odot}$ might be underestimated.
The observed velocity dispersion may be much smaller by chance, or J0248 could be an outlier in the BH-galaxy-halo relations, but these hypotheses can not be tested without more data on individual quasars.

The right panel of Figure \ref{fig:mhalo_ms_mhalo} shows the stellar mass-halo mass relation. We note that the host galaxy stellar mass and halo mass are derived independently, the former from the $M_{\rm BH}-M_{\rm \star, bulge}$ relation and the latter from group velocity dispersion. We compare our sample to the stellar mass-halo mass relation relations from direct measurements of low-redshift galaxy clusters \cite[][]{Kravtsov_etal_2018} and the abundance matching ansatz \citep[][]{Behroozi_etal_2013}. 
Across all halo mass ranges, the stellar mass-halo mass relation has relatively small scatter in stellar mass ($\approx0.2$\,dex) at fixed halo masses \citep{Tinker_etal_2013}.
While the \cite{Kravtsov_etal_2018} relation is derived for $z<0.1$ only, the stellar mass-halo mass relation is not expected to evolve much with redshift up to at least $z\approx4$ as shown in \cite{Behroozi_etal_2013, Behroozi_etal_2019}. 
Despite large uncertainties in the stellar masses inferred from the $M_{\rm BH}-M_{\rm \star, bulge}$ relation (e.g., $\approx0.5$\,dex considering measurement uncertainty and intrinsic scatter), most of our sample lies above the mean stellar mass-halo mass relation from abundance matching, hinting that luminous quasar hosts may represent outliers in the $M_{\rm BH}-M_{\star}$ and $M_{\rm BH}-M_{\rm h}$ relations or have less reliable virial BH masses.

\begin{figure*}
\centering
    \includegraphics[width=0.32\textwidth]{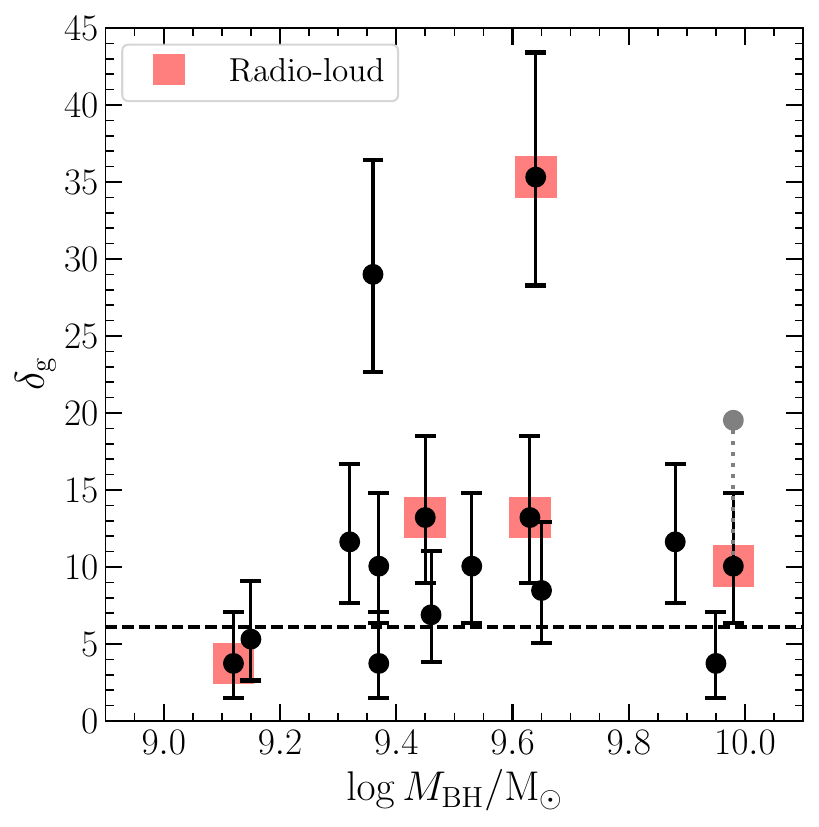}
    \includegraphics[width=0.32\textwidth]{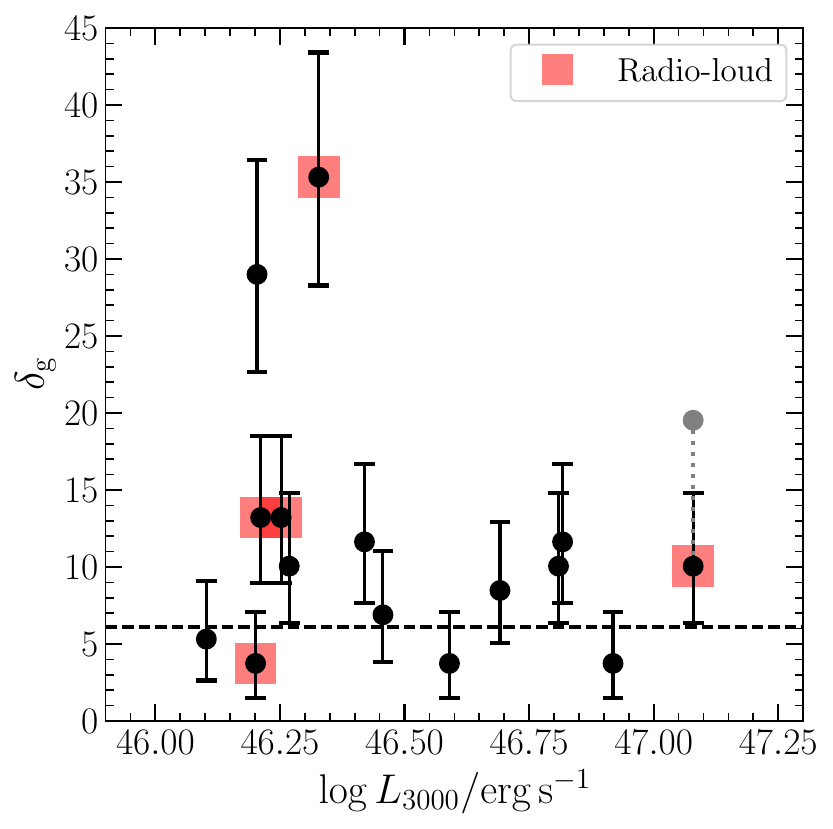}
    \includegraphics[width=0.32\textwidth]{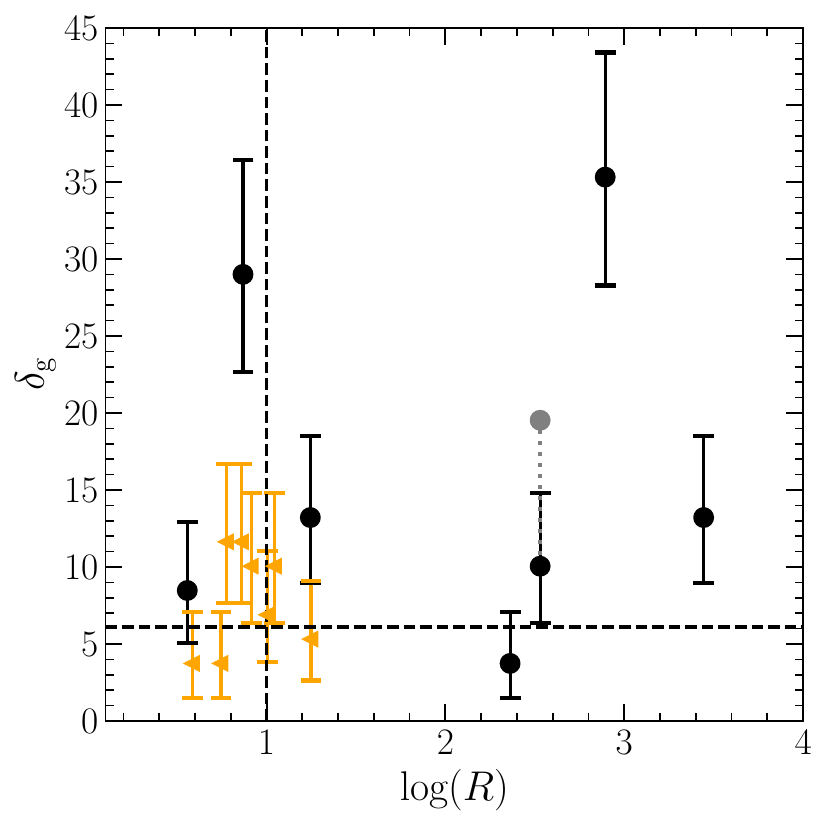}
\caption{Galaxy overdensity as a function of BH mass (left), bolometric luminosity (middle), and radio loudness (right). 
The orange triangle shows the 3$\sigma$ upper limit of the radio-loudness. The vertical and horizontal dashed lines indicate the $2\sigma$ overdensity and the radio-loud limit. We do not see a correlation between overdensity and BH mass, quasar luminosity, or radio-loudness in our sample.}
\label{fig:od_quasar}
\end{figure*}

\subsection{Overdensity and its dependency on quasar properties} 

At $z<0.5$, \cite{Serber_etal_2006} and \cite{Karhunen_etal_2014} found that quasars reside in denser environments at hundreds of kpc scale, but the overdensity vanishes at $\approx1$\, Mpc scale compared to galaxies with comparable masses. However, using spectroscopic redshifts, \cite{Wethers_etal_2022} found quasars tend to reside in moderate groups, and there is no statistical difference compared to the redshift- and stellar mass-matched control galaxy sample at the sub-Mpc scale. At $1<z<3$, the focus of quasar environment studies shifts to identifying protoclusters using radio-loud galaxies and AGN. For example, \cite{Wylezalek_etal_2013} found that the majority of radio-loud AGN resides in richer environment than average, with 55\% (10\%) of their sample in overdensities at $>2\sigma$ ($>5\sigma$) level on the Mpc-scale. In a redshift survey with HST WFC3 grism, \cite{Stott_etal_2020} found an overdensity in 8 out of 12 quasar fields in their sample, with a stacked overdensity of $\approx6\sigma$. Similarly, \cite{Trainor_etal_2012} revealed a significant stacked overdensity on the scale of 200\,kpc around the most luminous quasars at $z\simeq2.7$ in the Keck Baryonic Structure Survey (KBSS). In this work, we find that the majority (11 out of 15) of CUBS quasars reside in $>2\sigma$ overdensities within a projected distance of 250\,kpc (or $<1$\, Mpc including the LDSS3/IMACS FOV) and has a stacked overdensity of $\approx11\sigma$, consistent with the overdensity at both lower and higher redshifts in the literature. The median group size is 7 galaxies, roughly $\approx2\sigma$ overdensity above the background. In addition, we show that there is substantial diversity in the environments in which quasars reside. In this section, we explore the dependency of overdensity on various quasar and group properties.

Figure \ref{fig:od_quasar} shows the overdensity as a function of the SMBH mass (left panel) and quasar luminosity (middle panel). If the BH mass is a good tracer of the halo mass, as previously mentioned in Section \ref{sec:bh-gal-halo}, we expect a correlation between the BH mass and overdensity. However, possibly due to the limited dynamical range and large uncertainties associated with the virial BH mass and halo mass, we do not see a correlation between overdensity and BH mass in our sample.
The Pearson correlation coefficients are 0.21 and $-$0.13, with p-values of 0.46 and 0.65, for the $M_{\rm BH}-\delta_{g}$ and $L_{\rm 3000}-\delta_{g}$ relations, respectively.
There is no consensus on whether quasar environments depend on luminosity and BH mass in the literature. \cite{Karhunen_etal_2014} find that galaxy number density within 1\, Mpc has no dependency on redshift, quasar and host galaxy luminosity, or BH mass for low-redshift ($z<0.5$) quasars, and \cite{Zhang_etal_2013} find that clustering amplitude does not depend on redshift, luminosity, or BH mass, in SDSS Stripe 82 quasars over a wide redshift range of $0.6<z<1.2$. On the other hand, while there is no luminosity trend for the entire sample, \cite{Shen_etal_2009} found that the brightest 10\% of quasars are more strongly clustered than the remaining 90\% of quasars, suggesting only the brightest quasars reside in the rarest, most massive groups. Similarly, \cite{Shen_etal_2013} finds a weak luminosity dependency in quasar analysis at $z\approx0.5$. The CUBS quasars are no doubt some of the brightest quasars at $z\approx1$ (see Figure \ref{fig:quasar}), and while the overall stacked overdensity is significant, some CUBS quasars remain in moderate-to-poor environments (i.e., no overdensity compared to the background). 

Many studies have shown that radio-loud AGN/quasars tend to reside in denser regions than their radio-quiet counterparts and radio galaxies \citep[e.g.,][]{Wylezalek_etal_2013, RamosAlmeida_etal_2013, Hatch_etal_2014, Stott_etal_2020}, but others have found no significant difference \citep[e.g.,][]{Mclure_Dunlop_2001, Karhunen_etal_2014}. Figure \ref{fig:od_quasar} (right panel) shows the majority (4/5) of our radio-loud quasars reside in regions of high overdensity, while only roughly half (6/10) of the radio-quiet quasars show $>2\sigma$ overdensity. Similar to the literature finding radio-loud AGN in overdense environments, the connection between radio-loudness and overdensity is not straightforward and has significant scatter. In the simple AGN feedback evolution model from \cite{Hickox_etal_2009}, the central SMBH/galaxy grows mass from an optical/IR bright phase until reaching a critical halo mass around $10^{12}-10^{13} {\rm M}_{\odot}$, then shifts into a more quiescent phase with intermittent radio activity, which could be triggered by mergers and interactions. In this scenario, denser environments could lead to more mergers and interactions, and thus radio-loud quasars preferentially reside in overdense regions.

\section{Conclusions}\label{sec:conclusions}

Using a MUSE redshift survey and photometric follow-up observations, we study the group environments of the 15 UV-luminous quasars from the CUBS survey. We first identify galaxy group members around the CUBS quasars and measure the group properties and stellar properties of individual member galaxies. 

We find that the CUBS quasars reside in groups of sizes ranging from 3--26 galaxies, with halo masses of $10^{13}-10^{14}\,{\rm M}_{\odot}$. 
Out of 15 CUBS quasars, 11 are located in environments with overdensities greater than $2\sigma$ relative to the background (i.e., without the presence of a quasar). The overdensity of the quasar environment does not correlate with BH properties (BH mass, bolometric luminosity, and radio-loudness), but there is a tendency for radio-loud quasars to reside in overdense regions. Despite large uncertainties and scatter, the CUBS quasars deviate from the average stellar mass-halo mass relation, suggesting they could be outliers in the BH-galaxy-halo relations or have less reliable BH masses.

This study shows that quasars reside in diverse environments and that the relations between quasars and their environments are complicated. More deep, spectroscopic surveys around {\it individual} quasars spanning wider luminosity and redshift ranges with wide-field IFS like MUSE, are crucial to quantify the relations, as well as the intrinsic scatter in these relations, between quasars and their environments.

\begin{table*}
\movetableright=-1in
\caption{Summary of galaxy properties in each quasar field}
\begin{tabular}{rcccrrrrrrrr}
\hline\hline

 Quasar Name/ &       R. A. &          Decl. &      $z$ & ${m}_{\rm r}$ & ${M}_{\rm B}$ &  $B-I$ & $\log M_{\rm *}/M_{\odot}$ & $\Delta\,v$ & $\Delta\,\theta$ & $\Delta\,d$ \\
 
 Galaxy ID &  (J2000)     &   (J2000)     &    & (AB)   &  (AB) &  (AB) &  & (${\rm km}\,{\rm s}^{-1}$) & (arcsec) & (kpc) \\
  
\hline
J0028 &&&&&&&&&&\\  1 &  00:28:30.89 &  $-$33:05:55.89 &  0.8920 &        23.0 &     $-$18.6 &           1.8 &  ${10.05}^{-0.16}_{+0.16}$ &    $+$747 &            9.8 &       76 \\
  2 &  00:28:29.51 &  $-$33:05:52.16 &  0.8898 &        23.9 &     $-$17.9 &           1.8 &   ${9.45}^{-0.05}_{+0.05}$ &    $+$397 &           13.7 &      106 \\
  3 &  00:28:31.20 &  $-$33:06:14.13 &  0.8916 &        23.1 &     $-$18.1 &           0.7 &   ${8.86}^{-0.32}_{+0.18}$ &    $+$683 &           27.6 &      214 \\
  4 &  00:28:29.45 &  $-$33:06:17.18 &  0.8929 &        22.9 &     $-$19.0 &           1.6 &  ${10.05}^{-0.07}_{+0.07}$ &    $+$890 &           31.4 &      244 \\
  5 &  00:28:30.01 &  $-$33:06:20.44 &  0.8898 &        20.9 &     $-$21.2 &           1.8 &  ${10.97}^{-0.06}_{+0.04}$ &    $+$397 &           31.8 &      246 \\
  6 &  00:28:28.14 &  $-$33:05:38.58 &  0.8909 &        23.9 &     $-$17.3 &           0.9 &   ${9.51}^{-0.16}_{+0.20}$ &    $+$572 &           35.6 &      276 \\
\hline
J0110 &&&&&&&&&&\\
  1 &  01:10:35.52 &  $-$16:48:13.60 &  0.7840 &        23.3 &     $-$17.5 &           1.5 &  ${8.55}^{-0.15}_{+0.16}$ &    $+$286 &           14.1 &      105 \\
  2 &  01:10:35.73 &  $-$16:48:54.70 &  0.7839 &        23.9 &     $-$17.3 &           2.2 &  ${9.40}^{-0.15}_{+0.20}$ &    $+$269 &           27.2 &      203 \\

\hline
J0111 &&&&&&&&&&\\
  1 &  01:11:39.46 &  $-$03:15:39.54 &  1.2396 &        22.5 &     $-$19.7 &           0.6 &   ${9.51}^{-0.09}_{+0.09}$ &    $+$161 &           31.6 &      264 \\
  2 &  01:11:39.65 &  $-$03:15:36.30 &  1.2389 &        22.6 &     $-$21.0 &           1.6 &  ${10.63}^{-0.05}_{+0.06}$ &     $+$60 &           35.3 &      294 \\
\hline
J0114 &&&&&&&&&&\\
  1 &  01:14:22.19 &  $-$41:29:51.07 &  1.0236 &        23.3 &     $-$19.5 &           2.0 &  ${10.58}^{-0.09}_{+0.09}$ &     $-$30 &            3.9 &       31 \\
  2 &  01:14:21.42 &  $-$41:29:41.37 &  1.0274 &        23.4 &     $-$19.2 &           1.6 &   ${9.98}^{-0.06}_{+0.06}$ &    $+$533 &           12.1 &       97 \\
  3 &  01:14:20.81 &  $-$41:29:38.42 &  1.0241 &        24.3 &     $-$18.0 &           1.8 &   ${9.61}^{-0.28}_{+0.29}$ &     $+$44 &           21.5 &      174 \\
  4 &  01:14:20.57 &  $-$41:29:44.50 &  1.0228 &        21.4 &     $-$21.1 &           1.9 &  ${11.02}^{-0.03}_{+0.04}$ &    $-$148 &           23.4 &      188 \\
  5 &  01:14:20.12 &  $-$41:29:45.25 &  1.0187 &        21.5 &     $-$21.1 &           2.2 &  ${11.05}^{-0.06}_{+0.06}$ &    $-$755 &           30.2 &      243 \\
  6 &  01:14:20.04 &  $-$41:29:46.24 &  1.0243 &        23.1 &     $-$19.5 &           2.6 &  ${10.76}^{-0.06}_{+0.05}$ &     $+$74 &           31.3 &      252 \\
  7 &  01:14:20.93 &  $-$41:29:14.05 &  1.0285 &        22.1 &     $-$19.6 &           0.8 &   ${9.67}^{-0.10}_{+0.11}$ &    $+$696 &           37.7 &      304 \\
  8 &  01:14:19.70 &  $-$41:29:57.59 &  1.0222 &        23.5 &     $-$18.1 &           0.4 &   ${9.05}^{-0.22}_{+0.18}$ &    $-$237 &           37.8 &      304 \\
\hline
J0119 &&&&&&&&&&\\
  1 &  01:19:56.62 &  $-$20:10:22.46 &  0.8148 &        20.8 &     $-$19.4 &           0.8 &  ${9.54}^{-0.08}_{+0.07}$ &    $-$248 &            8.0 &       60 \\
  2 &  01:19:57.44 &  $-$20:10:29.75 &  0.8210 &        22.3 &     $-$18.6 &           1.0 &  ${9.38}^{-0.11}_{+0.10}$ &    $+$776 &           21.4 &      162 \\
  3 &  01:19:56.20 &  $-$20:10:46.06 &  0.8175 &        24.5 &     $-$16.3 &           1.0 &  ${8.39}^{-0.23}_{+0.24}$ &    $+$198 &           23.4 &      177 \\
  4 &  01:19:54.43 &  $-$20:10:34.28 &  0.8166 &        21.4 &     $-$19.6 &           1.2 &  ${9.79}^{-0.05}_{+0.07}$ &     $+$50 &           27.5 &      208 \\
  5 &  01:19:57.90 &  $-$20:10:00.94 &  0.8227 &        22.8 &     $-$18.2 &           1.3 &  ${9.27}^{-0.05}_{+0.05}$ &   $+$1056 &           34.8 &      263 \\
  6 &  01:19:57.58 &  $-$20:10:55.46 &  0.8223 &        21.9 &     $-$19.2 &           1.3 &  ${9.84}^{-0.07}_{+0.09}$ &    $+$990 &           39.6 &      299 \\
  7 &  01:19:53.71 &  $-$20:09:57.14 &  0.8204 &        20.3 &     $-$20.7 &           0.5 &  ${9.96}^{-0.07}_{+0.06}$ &    $+$677 &           44.0 &      332 \\
\hline
J0154 &&&&&&&&&&\\
  1 &  01:54:54.63 &  $-$07:11:55.52 &  1.2966 &        23.7 &     $-$19.3 &           1.2 &  ${10.49}^{-0.26}_{+0.23}$ &    $+$471 &           26.7 &      223 \\
  2 &  01:54:52.99 &  $-$07:12:13.70 &  1.2873 &        21.8 &     $-$21.1 &           1.4 &  ${10.45}^{-0.05}_{+0.06}$ &    $-$745 &           26.7 &      224 \\
  3 &  01:54:53.60 &  $-$07:11:55.28 &  1.2968 &        24.4 &     $-$19.4 &           2.7 &  ${10.24}^{-0.30}_{+0.32}$ &    $+$497 &           31.4 &      263 \\
  4 &  01:54:55.86 &  $-$07:12:49.74 &  1.2953 &        23.4 &     $-$20.3 &           2.0 &  ${10.69}^{-0.06}_{+0.05}$ &    $+$301 &           32.8 &      274 \\
  5 &  01:54:56.13 &  $-$07:12:50.45 &  1.2973 &        24.1 &     $-$18.2 &           0.8 &   ${8.50}^{-0.30}_{+0.36}$ &    $+$562 &           35.6 &      298 \\
  6 &  01:54:52.26 &  $-$07:12:38.91 &  1.2948 &        23.6 &     $-$19.5 &           1.8 &  ${10.31}^{-0.27}_{+0.23}$ &    $+$235 &           40.0 &      335 \\
\hline
J0248 &&&&&&&&&&\\
  1 &  02:48:06.70 &  $-$40:48:33.82 &  0.8845 &        20.7 &     $-$20.9 &           0.8 &  ${10.10}^{-0.02}_{+0.01}$ &     $+$16 &            6.2 &       48 \\
  2 &  02:48:06.87 &  $-$40:48:31.44 &  0.8834 &        24.4 &     $-$17.3 &           2.6 &  ${10.01}^{-0.20}_{+0.20}$ &    $-$159 &            9.0 &       70 \\
  3 &  02:48:07.27 &  $-$40:48:14.64 &  0.8835 &        24.1 &     $-$17.5 &           1.7 &   ${9.70}^{-0.15}_{+0.15}$ &    $-$151 &           24.1 &      187 \\
  4 &  02:48:03.85 &  $-$40:48:13.47 &  0.8853 &        20.7 &     $-$21.0 &           1.8 &  ${10.85}^{-0.06}_{+0.04}$ &    $+$143 &           41.8 &      324 \\
  5 &  02:48:04.07 &  $-$40:48:01.62 &  0.8839 &        22.2 &     $-$19.5 &           1.8 &  ${10.14}^{-0.05}_{+0.06}$ &     $-$80 &           46.2 &      358 \\
\hline\hline
\end{tabular}
\label{tab:galaxy}
 \tablecomments{The table columns are quasar name/galaxy ID, galaxy right ascension, declination, redshift, $r$-band magnitude from MUSE, $B$-band magnitude from SED fitting, $B-I$ color, stellar mass, relative velocity to the quasar redshift, angular distance and projected distance to the quasar.}
\end{table*}

\begin{table*}
\movetableright=-1in
\setcounter{table}{2}
\caption{(Continued)}
\begin{tabular}{rcccrrrrrrrr}
\hline
\hline
 Quasar Name/ &       R. A. &          Decl. &      $z$ & ${m}_{\rm r}$ & ${M}_{\rm B}$ &  $B-I$ & $\log M_{\rm *}/M_{\odot}$ & $\Delta\,v$ & $\Delta\,\theta$ & $\Delta\,d$ \\
 
 Galaxy ID &  (J2000)     &   (J2000)     &    & (AB)   &  (AB) &  (AB) &  & (${\rm km}\,{\rm s}^{-1}$) & (arcsec) & (kpc) \\
\hline
J0333 &&&&&&&&&&\\
  1 &  03:33:07.34 &  $-$41:02:00.99 &  1.1079 &        -- &     -- &          -- &  --$^{a}$ &   $-$1049 &            3.9 &       32 \\
  2 &  03:33:07.36 &  $-$41:01:59.39 &  1.1114 &        -- &     -- &          -- &  --$^{a}$ &    $-$553 &            4.7 &       38 \\
  3 &  03:33:05.93 &  $-$41:01:57.61 &  1.1104 &        24.1 &     $-$18.9 &           2.3 &  ${10.27}^{-0.20}_{+0.18}$ &    $-$694 &           17.5 &      143 \\
  4 &  03:33:06.64 &  $-$41:01:44.42 &  1.1126 &        23.2 &     $-$19.0 &           0.9 &   ${9.00}^{-0.23}_{+0.23}$ &    $-$383 &           17.9 &      147 \\
  5 &  03:33:05.72 &  $-$41:02:11.20 &  1.1139 &        21.9 &     $-$20.2 &           1.3 &   ${9.86}^{-0.17}_{+0.19}$ &    $-$198 &           22.7 &      186 \\
  6 &  03:33:08.77 &  $-$41:01:58.96 &  1.1121 &        23.4 &     $-$18.9 &           1.3 &   ${9.40}^{-0.19}_{+0.23}$ &    $-$454 &           25.5 &      209 \\
  7 &  03:33:10.16 &  $-$41:01:51.01 &  1.1124 &        21.0 &     $-$20.9 &           0.9 &   ${9.89}^{-0.11}_{+0.11}$ &    $-$411 &           47.3 &      388 \\
\hline
J0357 &&&&&&&&&&\\
  1 &  03:57:22.14 &  $-$48:12:16.18 &  1.0176 &        -- &     -- &           -- & --$^{a}$ &    $+$715 &            3.6 &       29 \\
  2 &  03:57:21.72 &  $-$48:12:12.58 &  1.0148 &        -- &     -- &           -- & --$^{a}$ &   $+$298 &            3.9 &       31 \\
  3 &  03:57:22.23 &  $-$48:12:17.79 &  1.0132 &        22.8 &     $-$19.9 &           1.8 &  ${10.65}^{-0.01}_{+0.02}$ &     $+$67 &            5.3 &       43 \\
  4 &  03:57:21.84 &  $-$48:12:26.52 &  1.0167 &        23.7 &     $-$18.2 &           0.8 &   ${9.32}^{-0.20}_{+0.17}$ &    $+$581 &           11.4 &       92 \\
  5 &  03:57:21.15 &  $-$48:12:05.54 &  1.0132 &        21.7 &     $-$20.5 &           1.7 &  ${10.48}^{-0.04}_{+0.05}$ &     $+$67 &           15.0 &      121 \\
  6 &  03:57:22.88 &  $-$48:12:22.55 &  1.0147 &        23.6 &     $-$18.8 &           1.4 &   ${9.47}^{-0.05}_{+0.05}$ &    $+$283 &           16.2 &      130 \\
  7 &  03:57:21.07 &  $-$48:12:04.39 &  1.0156 &        23.1 &     $-$19.4 &           1.1 &   ${9.68}^{-0.11}_{+0.12}$ &    $+$417 &           16.7 &      134 \\
  8 &  03:57:20.85 &  $-$48:12:23.18 &  1.0227 &        23.2 &     $-$19.0 &           1.7 &   ${9.55}^{-0.19}_{+0.24}$ &   $+$1475 &           17.9 &      144 \\
  9 &  03:57:20.92 &  $-$48:12:01.91 &  1.0156 &        22.4 &     $-$19.8 &           1.8 &  ${10.42}^{-0.06}_{+0.05}$ &    $+$417 &           20.0 &      160 \\
 10 &  03:57:21.93 &  $-$48:12:35.99 &  1.0241 &        22.5 &     $-$19.6 &           1.2 &   ${9.81}^{-0.19}_{+0.15}$ &   $+$1683 &           20.8 &      167 \\
 11 &  03:57:22.77 &  $-$48:11:58.69 &  1.0183 &        22.3 &     $-$19.6 &           0.7 &   ${9.44}^{-0.06}_{+0.07}$ &    $+$819 &           20.9 &      168 \\
 12 &  03:57:21.91 &  $-$48:12:37.02 &  1.0226 &        21.6 &     $-$20.8 &           2.6 &  ${11.55}^{-0.14}_{+0.15}$ &   $+$1460 &           21.9 &      176 \\
 13 &  03:57:23.45 &  $-$48:12:17.37 &  1.0113 &        22.9 &     $-$19.0 &           1.2 &   ${9.51}^{-0.14}_{+0.16}$ &    $-$223 &           23.1 &      186 \\
 14 &  03:57:23.47 &  $-$48:12:23.92 &  1.0186 &        21.9 &     $-$20.3 &           1.6 &  ${10.38}^{-0.06}_{+0.05}$ &    $+$864 &           24.9 &      200 \\
 15 &  03:57:20.32 &  $-$48:12:28.95 &  1.0168 &        21.4 &     $-$21.1 &           1.3 &  ${10.44}^{-0.06}_{+0.05}$ &    $+$596 &           27.6 &      222 \\
 16 &  03:57:23.43 &  $-$48:12:32.35 &  1.0118 &        21.9 &     $-$20.3 &           2.2 &  ${10.69}^{-0.03}_{+0.07}$ &    $-$149 &           28.5 &      229 \\
 17 &  03:57:23.25 &  $-$48:12:41.09 &  1.0097 &        22.8 &     $-$18.6 &           0.4 &   ${9.30}^{-0.13}_{+0.09}$ &    $-$462 &           32.8 &      263 \\
 18 &  03:57:24.08 &  $-$48:12:25.86 &  1.0171 &        22.3 &     $-$19.8 &           1.5 &  ${10.14}^{-0.05}_{+0.05}$ &    $+$633 &           34.2 &      274 \\
 19 &  03:57:20.17 &  $-$48:11:48.50 &  1.0143 &        23.0 &     $-$18.8 &           0.9 &   ${9.30}^{-0.17}_{+0.21}$ &    $+$223 &           37.4 &      300 \\
 20 &  03:57:25.00 &  $-$48:12:21.41 &  1.0109 &        22.0 &     $-$19.8 &           0.5 &   ${9.43}^{-0.06}_{+0.06}$ &    $-$283 &           46.7 &      375 \\
 21 &  03:57:24.53 &  $-$48:12:43.97 &  1.0135 &        21.5 &     $-$20.7 &           1.0 &  ${10.14}^{-0.06}_{+0.06}$ &    $+$104 &           48.7 &      391 \\
 22 &  03:57:25.30 &  $-$48:12:44.09 &  1.0106 &        22.3 &     $-$19.8 &           1.4 &   ${9.90}^{-0.14}_{+0.15}$ &    $-$328 &           58.4 &      469 \\

\hline
J0420 &&&&&&&&&&\\
  1 &  04:20:53.71 &  $-$56:50:42.45 &  0.9476 &       -- &     -- &           -- & --$^{a}$ &     $-$77 &            3.3 &       26 \\
  2 &  04:20:52.52 &  $-$56:50:47.40 &  0.9465 &        22.0 &     $-$19.3 &           1.5 &   ${9.67}^{-0.10}_{+0.15}$ &    $-$246 &           21.1 &      166 \\
  3 &  04:20:54.79 &  $-$56:51:01.53 &  0.9495 &        23.3 &     $-$18.0 &           0.7 &   ${8.50}^{-0.16}_{+0.16}$ &    $+$215 &           22.0 &      174 \\
\hline
J0454 &&&&&&&&&&\\
  1 &  04:54:16.24 &  $-$61:16:28.79 &  0.7879 &        -- &     -- &           -- & --$^{a}$ &   $+$302 &            4.8 &       36 \\
  2 &  04:54:15.89 &  $-$61:16:32.48 &  0.7867 &        22.6 &     $-$18.4 &           1.4 &   ${9.48}^{-0.14}_{+0.14}$ &    $+$101 &            6.0 &       45 \\
  3 &  04:54:15.05 &  $-$61:16:36.04 &  0.7890 &        21.9 &     $-$19.1 &           1.6 &   ${9.89}^{-0.07}_{+0.07}$ &    $+$487 &           16.6 &      124 \\
  4 &  04:54:15.10 &  $-$61:16:14.03 &  0.7878 &        19.0 &     $-$22.4 &           1.3 &  ${11.06}^{-0.04}_{+0.04}$ &    $+$285 &           17.9 &      134 \\
  5 &  04:54:14.77 &  $-$61:16:19.83 &  0.7853 &        21.0 &     $-$19.9 &           1.5 &  ${10.13}^{-0.04}_{+0.06}$ &    $-$134 &           19.0 &      142 \\
  6 &  04:54:16.53 &  $-$61:16:09.50 &  0.7890 &        20.4 &     $-$20.9 &           1.8 &  ${10.82}^{-0.03}_{+0.04}$ &    $+$487 &           19.1 &      143 \\
  7 &  04:54:15.85 &  $-$61:16:46.43 &  0.7880 &        22.8 &     $-$17.7 &           2.2 &   ${9.61}^{-0.23}_{+0.25}$ &    $+$319 &           19.9 &      149 \\
  8 &  04:54:17.45 &  $-$61:16:32.66 &  0.7869 &        20.2 &     $-$21.2 &           1.6 &  ${10.92}^{-0.04}_{+0.03}$ &    $+$134 &           23.2 &      173 \\
  9 &  04:54:17.12 &  $-$61:16:42.52 &  0.7858 &        22.6 &     $-$18.4 &           1.5 &   ${9.57}^{-0.06}_{+0.07}$ &     $-$50 &           23.7 &      177 \\
 10 &  04:54:16.98 &  $-$61:16:52.17 &  0.7868 &        23.0 &     $-$18.1 &           1.0 &   ${9.02}^{-0.09}_{+0.08}$ &    $+$117 &           29.9 &      223 \\
 11 &  04:54:16.77 &  $-$61:15:57.71 &  0.7867 &        22.1 &     $-$19.2 &           1.8 &  ${10.07}^{-0.06}_{+0.07}$ &    $+$101 &           31.4 &      234 \\
 12 &  04:54:18.43 &  $-$61:16:37.22 &  0.7845 &        23.1 &     $-$17.5 &           0.7 &   ${8.39}^{-0.12}_{+0.13}$ &    $-$269 &           38.7 &      289 \\

\hline\hline
\end{tabular}
 \tablecomments{$^{a}$ failed to decompose galaxy and quasar light with PSF subtraction}
\end{table*}

\begin{table*}
\movetableright=-1in
\setcounter{table}{2}
\caption{(Continued)}
\begin{tabular}{rcccrrrrrrrr}
\hline\hline
 Quasar Name/ &       R. A. &          Decl. &      $z$ & ${m}_{\rm r}$ & ${M}_{\rm B}$ &  $B-I$ & $\log M_{\rm *}/M_{\odot}$ & $\Delta\,v$ & $\Delta\,\theta$ & $\Delta\,d$ \\
 
Galaxy ID  &  (J2000)     &   (J2000)     &    & (AB)   &  (AB) &  (AB) &  & (${\rm km}\,{\rm s}^{-1}$) & (arcsec) & (kpc) \\
\hline
 13 &  04:54:12.81 &  $-$61:16:27.45 &  0.7867 &        23.1 &     $-$17.6 &           0.9 &   ${9.19}^{-0.17}_{+0.17}$ &    $+$101 &           47.1 &      351 \\
 14 &  04:54:19.32 &  $-$61:16:38.62 &  0.7887 &        21.8 &     $-$19.1 &           1.8 &  ${10.11}^{-0.06}_{+0.06}$ &    $+$436 &           52.0 &      388 \\
 15 &  04:54:12.09 &  $-$61:16:30.84 &  0.7890 &        22.9 &     $-$18.2 &           1.4 &   ${9.47}^{-0.07}_{+0.06}$ &    $+$487 &           58.1 &      434 \\
 16 &  04:54:11.84 &  $-$61:16:34.53 &  0.7890 &        23.6 &     $-$17.3 &           1.6 &   ${9.48}^{-0.09}_{+0.09}$ &    $+$487 &           62.2 &      464 \\
 17 &  04:54:19.98 &  $-$61:16:07.68 &  0.7878 &       -99.0 &     $-$21.8 &           1.7 &  ${11.17}^{-0.05}_{+0.05}$ &    $+$285 &           63.4 &      473 \\
 18 &  04:54:11.90 &  $-$61:16:08.13 &  0.7820 &        23.6 &     $-$17.0 &           1.2 &   ${8.96}^{-0.17}_{+0.16}$ &    $-$688 &           63.5 &      474 \\
 
\hline
J2135 &&&&&&&&&&\\
  1 &  21:35:53.60 &  $-$53:16:49.30 &  0.8125 &         -- &     -- &          -- &  --$^{b}$ &    $+$165 &            8.9 &       67 \\
  2 &  21:35:51.60 &  $-$53:16:53.63 &  0.8133 &        22.7 &     $-$18.3 &           1.5 &  ${9.55}^{-0.16}_{+0.16}$ &    $+$298 &           24.1 &      182 \\
\hline
J2245 &&&&&&&&&&\\
  1 &  22:45:00.50 &  $-$49:31:34.19 &  1.0023 &        20.6 &     $-$21.7 &           1.6 &  ${10.91}^{-0.05}_{+0.04}$ &    $+$180 &           14.9 &      120 \\
  2 &  22:45:01.29 &  $-$49:31:49.00 &  1.0034 &        22.6 &     $-$20.2 &           2.2 &  ${10.73}^{-0.04}_{+0.05}$ &    $+$345 &           16.2 &      130 \\
  3 &  22:44:58.92 &  $-$49:31:38.39 &  1.0007 &        23.4 &     $-$19.1 &           2.0 &  ${10.33}^{-0.05}_{+0.06}$ &     $-$60 &           21.8 &      175 \\
  4 &  22:45:01.63 &  $-$49:31:55.05 &  1.0048 &        22.1 &     $-$20.5 &           2.0 &  ${10.77}^{-0.05}_{+0.04}$ &    $+$554 &           22.3 &      179 \\
  5 &  22:45:00.12 &  $-$49:31:17.95 &  1.0036 &        24.0 &     $-$17.5 &           0.9 &   ${8.75}^{-0.36}_{+0.29}$ &    $+$370 &           30.5 &      245 \\
  6 &  22:45:01.47 &  $-$49:31:19.30 &  1.0027 &        21.9 &     $-$19.4 &           1.1 &   ${9.89}^{-0.11}_{+0.08}$ &    $+$240 &           34.8 &      279 \\
  7 &  22:45:00.86 &  $-$49:31:15.03 &  1.0003 &        21.8 &     $-$20.4 &           1.8 &  ${10.71}^{-0.05}_{+0.06}$ &    $-$120 &           34.8 &      279 \\
  8 &  22:44:57.19 &  $-$49:31:32.88 &  0.9996 &        22.5 &     $-$19.0 &           1.1 &   ${9.08}^{-0.19}_{+0.21}$ &    $-$225 &           47.8 &      383 \\
\hline
J2308 &&&&&&&&&&\\
  1 &  23:08:37.87 &  $-$52:58:46.21 &  1.0760 &        -- &    -- &         -- &  --$^{a}$ &    $+$390 &            2.9 &       24 \\
  2 &  23:08:39.34 &  $-$52:58:37.37 &  1.0739 &        25.0 &     $-$17.3 &           0.8 &  ${8.72}^{-0.26}_{+0.32}$ &     $+$80 &           25.9 &      210 \\
  3 &  23:08:35.85 &  $-$52:58:50.78 &  1.0749 &        24.6 &     $-$17.3 &           0.6 &  ${8.80}^{-0.21}_{+0.30}$ &    $+$231 &           29.3 &      238 \\
  4 &  23:08:36.46 &  $-$52:59:16.87 &  1.0749 &        22.2 &     $-$19.6 &           0.8 &  ${9.73}^{-0.11}_{+0.09}$ &    $+$231 &           34.4 &      280 \\
\hline
J2339 &&&&&&&&&&\\
  1 &  23:39:13.35 &  $-$55:23:35.39 &  1.3482 &        24.4 &     $-$18.7 &           1.8 &  ${10.28}^{-0.24}_{+0.26}$ &    $-$789 &           15.6 &      131 \\
  2 &  23:39:12.14 &  $-$55:24:02.43 &  1.3606 &        23.7 &     $-$18.7 &           1.1 &   ${8.66}^{-0.37}_{+0.30}$ &    $+$789 &           20.0 &      168 \\
  3 &  23:39:11.93 &  $-$55:23:40.81 &  1.3560 &        21.5 &     $-$21.9 &           1.6 &  ${10.80}^{-0.05}_{+0.05}$ &    $+$210 &           21.8 &      184 \\
  4 &  23:39:14.51 &  $-$55:24:12.34 &  1.3575 &        23.5 &     $-$19.0 &           0.9 &   ${9.99}^{-0.19}_{+0.18}$ &    $+$395 &           29.0 &      244 \\
  5 &  23:39:15.46 &  $-$55:23:35.59 &  1.3625 &        24.0 &     $-$19.6 &           2.5 &  ${10.58}^{-0.12}_{+0.12}$ &   $+$1031 &           36.9 &      310 \\
  6 &  23:39:11.55 &  $-$55:23:21.51 &  1.3430 &        22.3 &     $-$21.6 &           2.1 &  ${11.10}^{-0.06}_{+0.06}$ &   $-$1452 &           38.5 &      324 \\
  7 &  23:39:15.35 &  $-$55:24:17.09 &  1.3426 &        23.1 &     $-$19.8 &           1.0 &   ${9.57}^{-0.22}_{+0.25}$ &   $-$1503 &           41.4 &      348 \\
  8 &  23:39:15.42 &  $-$55:24:19.02 &  1.3479 &        21.7 &     $-$21.1 &           1.1 &  ${10.01}^{-0.19}_{+0.16}$ &    $-$828 &           43.5 &      366 \\
  9 &  23:39:10.47 &  $-$55:23:32.67 &  1.3423 &        21.9 &     $-$21.0 &           1.8 &  ${10.76}^{-0.14}_{+0.11}$ &   $-$1534 &           45.1 &      379 \\
 10 &  23:39:10.00 &  $-$55:23:50.53 &  1.3624 &        23.3 &     $-$20.1 &           1.4 &   ${9.65}^{-0.26}_{+0.23}$ &   $+$1019 &           48.3 &      406 \\
 11 &  23:39:16.29 &  $-$55:23:25.81 &  1.3477 &        24.1 &     $-$18.9 &           1.2 &   ${9.47}^{-0.22}_{+0.28}$ &    $-$853 &           52.4 &      441 \\
 12 &  23:39:18.08 &  $-$55:24:16.86 &  1.3599 &        24.6 &     $-$17.8 &           0.7 &   ${9.04}^{-0.41}_{+0.38}$ &    $+$700 &           77.5 &      652 \\
\hline\hline
\end{tabular}
\tablecomments{$^{b}$ too faint to obtain pseudo-photometry from MUSE}
\end{table*}

\begin{acknowledgments}
{We thank the anonymous referee for useful comments that helped improve this work.}
JIL is supported by the Eric and Wendy Schmidt AI in Science Postdoctoral Fellowship, a Schmidt Futures program. 
EB acknowledges support by NASA under award number 80GSFC21M0002.
HWC and MCC acknowledge partial support from NSF AST-1715692 grants.
FSZ acknowledges support of a Carnegie Fellowship from the Observatories of the Carnegie Institution for Science.
Based on observations collected at the European Organisation for Astronomical Research in the Southern Hemisphere under ESO program 104.A-0147. This paper includes data gathered with the 6.5-meter Magellan Telescopes located at Las Campanas Observatory, Chile. 
\end{acknowledgments}

\bibliography{refs} 

\begin{thebibliography}{}
\expandafter\ifx\csname natexlab\endcsname\relax\def\natexlab#1{#1}\fi
\providecommand{\url}[1]{\href{#1}{#1}}
\providecommand{\dodoi}[1]{doi:~\href{http://doi.org/#1}{\nolinkurl{#1}}}
\providecommand{\doeprint}[1]{\href{http://ascl.net/#1}{\nolinkurl{http://ascl.net/#1}}}
\providecommand{\doarXiv}[1]{\href{https://arxiv.org/abs/#1}{\nolinkurl{https://arxiv.org/abs/#1}}}

\bibitem[{{Bacon} {et~al.}(2016){Bacon}, {Piqueras}, {Conseil}, {Richard}, \&
  {Shepherd}}]{Mpdaf}
{Bacon}, R., {Piqueras}, L., {Conseil}, S., {Richard}, J., \& {Shepherd}, M.
  2016, {MPDAF: MUSE Python Data Analysis Framework}, Astrophysics Source Code
  Library, record ascl:1611.003.
\newblock \doeprint{1611.003}

\bibitem[{{Bacon} {et~al.}(2010){Bacon}, {Accardo}, {Adjali}, {Anwand},
  {Bauer}, {Biswas}, {Blaizot}, {Boudon}, {Brau-Nogue}, {Brinchmann},
  {Caillier}, {Capoani}, {Carollo}, {Contini}, {Couderc}, {Daguis{\'e}},
  {Deiries}, {Delabre}, {Dreizler}, {Dubois}, {Dupieux}, {Dupuy}, {Emsellem},
  {Fechner}, {Fleischmann}, {Fran{\c{c}}ois}, {Gallou}, {Gharsa}, {Glindemann},
  {Gojak}, {Guiderdoni}, {Hansali}, {Hahn}, {Jarno}, {Kelz}, {Koehler},
  {Kosmalski}, {Laurent}, {Le Floch}, {Lilly}, {Lizon}, {Loupias}, {Manescau},
  {Monstein}, {Nicklas}, {Olaya}, {Pares}, {Pasquini}, {P{\'e}contal-Rousset},
  {Pell{\'o}}, {Petit}, {Popow}, {Reiss}, {Remillieux}, {Renault}, {Roth},
  {Rupprecht}, {Serre}, {Schaye}, {Soucail}, {Steinmetz}, {Streicher}, {Stuik},
  {Valentin}, {Vernet}, {Weilbacher}, {Wisotzki}, \& {Yerle}}]{MUSE}
{Bacon}, R., {Accardo}, M., {Adjali}, L., {et~al.} 2010, in Society of
  Photo-Optical Instrumentation Engineers (SPIE) Conference Series, Vol. 7735,
  Ground-based and Airborne Instrumentation for Astronomy III, ed. I.~S.
  {McLean}, S.~K. {Ramsay}, \& H.~{Takami}, 773508, \dodoi{10.1117/12.856027}

\bibitem[{{Behroozi} {et~al.}(2019){Behroozi}, {Wechsler}, {Hearin}, \&
  {Conroy}}]{Behroozi_etal_2019}
{Behroozi}, P., {Wechsler}, R.~H., {Hearin}, A.~P., \& {Conroy}, C. 2019,
  \mnras, 488, 3143, \dodoi{10.1093/mnras/stz1182}

\bibitem[{{Behroozi} {et~al.}(2013){Behroozi}, {Wechsler}, \&
  {Conroy}}]{Behroozi_etal_2013}
{Behroozi}, P.~S., {Wechsler}, R.~H., \& {Conroy}, C. 2013, \apj, 770, 57,
  \dodoi{10.1088/0004-637X/770/1/57}

\bibitem[{{Bennert} {et~al.}(2011){Bennert}, {Auger}, {Treu}, {Woo}, \&
  {Malkan}}]{Bennert_etal_2011}
{Bennert}, V.~N., {Auger}, M.~W., {Treu}, T., {Woo}, J.-H., \& {Malkan}, M.~A.
  2011, \apj, 742, 107, \dodoi{10.1088/0004-637X/742/2/107}

\bibitem[{{Bentz} {et~al.}(2013){Bentz}, {Denney}, {Grier}, {Barth},
  {Peterson}, {Vestergaard}, {Bennert}, {Canalizo}, {De Rosa}, {Filippenko},
  {Gates}, {Greene}, {Li}, {Malkan}, {Pogge}, {Stern}, {Treu}, \&
  {Woo}}]{Bentz_etal_2013}
{Bentz}, M.~C., {Denney}, K.~D., {Grier}, C.~J., {et~al.} 2013, \apj, 767, 149,
  \dodoi{10.1088/0004-637X/767/2/149}

\bibitem[{{Bertin} \& {Arnouts}(1996)}]{SourceExtractor}
{Bertin}, E., \& {Arnouts}, S. 1996, \aaps, 117, 393,
  \dodoi{10.1051/aas:1996164}

\bibitem[{{Booth} \& {Schaye}(2010)}]{Booth_Schaye_2010}
{Booth}, C.~M., \& {Schaye}, J. 2010, \mnras, 405, L1,
  \dodoi{10.1111/j.1745-3933.2010.00832.x}

\bibitem[{{Bruzual} \& {Charlot}(2003)}]{BC03}
{Bruzual}, G., \& {Charlot}, S. 2003, \mnras, 344, 1000,
  \dodoi{10.1046/j.1365-8711.2003.06897.x}

\bibitem[{{Calzetti} {et~al.}(2000){Calzetti}, {Armus}, {Bohlin}, {Kinney},
  {Koornneef}, \& {Storchi-Bergmann}}]{Calzetti_etal_2000}
{Calzetti}, D., {Armus}, L., {Bohlin}, R.~C., {et~al.} 2000, \apj, 533, 682,
  \dodoi{10.1086/308692}

\bibitem[{{Cantalupo} {et~al.}(2019){Cantalupo}, {Pezzulli}, {Lilly}, {Marino},
  {Gallego}, {Schaye}, {Bacon}, {Feltre}, {Kollatschny}, {Nanayakkara},
  {Richard}, {Wendt}, {Wisotzki}, \& {Prochaska}}]{Cantalupo_etal_2019}
{Cantalupo}, S., {Pezzulli}, G., {Lilly}, S.~J., {et~al.} 2019, \mnras, 483,
  5188, \dodoi{10.1093/mnras/sty3481}

\bibitem[{{Carnall} {et~al.}(2018){Carnall}, {McLure}, {Dunlop}, \&
  {Dav{\'e}}}]{Bagpipes1}
{Carnall}, A.~C., {McLure}, R.~J., {Dunlop}, J.~S., \& {Dav{\'e}}, R. 2018,
  \mnras, 480, 4379, \dodoi{10.1093/mnras/sty2169}

\bibitem[{{Carnall} {et~al.}(2019){Carnall}, {McLure}, {Dunlop}, {Cullen},
  {McLeod}, {Wild}, {Johnson}, {Appleby}, {Dav{\'e}}, {Amorin}, {Bolzonella},
  {Castellano}, {Cimatti}, {Cucciati}, {Gargiulo}, {Garilli}, {Marchi},
  {Pentericci}, {Pozzetti}, {Schreiber}, {Talia}, \& {Zamorani}}]{Bagpipes2}
{Carnall}, A.~C., {McLure}, R.~J., {Dunlop}, J.~S., {et~al.} 2019, \mnras, 490,
  417, \dodoi{10.1093/mnras/stz2544}

\bibitem[{{Charlot} \& {Fall}(2000)}]{Charlot_Fall_2000}
{Charlot}, S., \& {Fall}, S.~M. 2000, \apj, 539, 718, \dodoi{10.1086/309250}

\bibitem[{{Chen} {et~al.}(2020){Chen}, {Zahedy}, {Boettcher}, {Cooper},
  {Johnson}, {Rudie}, {Chen}, {Walth}, {Cantalupo}, {Cooksey},
  {Faucher-Gigu{\`e}re}, {Greene}, {Lopez}, {Mulchaey}, {Penton}, {Petitjean},
  {Putman}, {Rafelski}, {Rauch}, {Schaye}, {Simcoe}, \&
  {Weiner}}]{Chen_etal_2020_CUBS1}
{Chen}, H.-W., {Zahedy}, F.~S., {Boettcher}, E., {et~al.} 2020, \mnras, 497,
  498, \dodoi{10.1093/mnras/staa1773}

\bibitem[{{Decarli} {et~al.}(2010){Decarli}, {Falomo}, {Treves}, {Labita},
  {Kotilainen}, \& {Scarpa}}]{Decarli_etal_2010}
{Decarli}, R., {Falomo}, R., {Treves}, A., {et~al.} 2010, \mnras, 402, 2453,
  \dodoi{10.1111/j.1365-2966.2009.16049.x}

\bibitem[{{Di Matteo} {et~al.}(2005){Di Matteo}, {Springel}, \&
  {Hernquist}}]{DiMatteo_etal_2005}
{Di Matteo}, T., {Springel}, V., \& {Hernquist}, L. 2005, \nat, 433, 604,
  \dodoi{10.1038/nature03335}

\bibitem[{{Donahue} \& {Voit}(2022)}]{Donahue_Voit_2022}
{Donahue}, M., \& {Voit}, G.~M. 2022, \physrep, 973, 1,
  \dodoi{10.1016/j.physrep.2022.04.005}

\bibitem[{{Dressler} {et~al.}(2011){Dressler}, {Bigelow}, {Hare}, {Sutin},
  {Thompson}, {Burley}, {Epps}, {Oemler}, {Bagish}, {Birk}, {Clardy},
  {Gunnels}, {Kelson}, {Shectman}, \& {Osip}}]{IMACS}
{Dressler}, A., {Bigelow}, B., {Hare}, T., {et~al.} 2011, \pasp, 123, 288,
  \dodoi{10.1086/658908}

\bibitem[{{Ferragamo} {et~al.}(2020){Ferragamo}, {Rubi{\~n}o-Mart{\'\i}n},
  {Betancort-Rijo}, {Munari}, {Sartoris}, \& {Barrena}}]{Ferragamo_etal_2020}
{Ferragamo}, A., {Rubi{\~n}o-Mart{\'\i}n}, J.~A., {Betancort-Rijo}, J.,
  {et~al.} 2020, \aap, 641, A41, \dodoi{10.1051/0004-6361/201834837}

\bibitem[{{Ferrarese}(2002)}]{Ferrarese_2002}
{Ferrarese}, L. 2002, \apj, 578, 90, \dodoi{10.1086/342308}

\bibitem[{{Ferrarese} \& {Merritt}(2000)}]{Ferrarese_Merritt_2000}
{Ferrarese}, L., \& {Merritt}, D. 2000, \apjl, 539, L9, \dodoi{10.1086/312838}

\bibitem[{{Gaspari} {et~al.}(2019){Gaspari}, {Eckert}, {Ettori}, {Tozzi},
  {Bassini}, {Rasia}, {Brighenti}, {Sun}, {Borgani}, {Johnson}, {Tremblay},
  {Stone}, {Temi}, {Yang}, {Tombesi}, \& {Cappi}}]{Gaspari_etal_2019}
{Gaspari}, M., {Eckert}, D., {Ettori}, S., {et~al.} 2019, \apj, 884, 169,
  \dodoi{10.3847/1538-4357/ab3c5d}

\bibitem[{{Gebhardt} {et~al.}(2000){Gebhardt}, {Bender}, {Bower}, {Dressler},
  {Faber}, {Filippenko}, {Green}, {Grillmair}, {Ho}, {Kormendy}, {Lauer},
  {Magorrian}, {Pinkney}, {Richstone}, \& {Tremaine}}]{Gebhardt_etal_2000}
{Gebhardt}, K., {Bender}, R., {Bower}, G., {et~al.} 2000, \apjl, 539, L13,
  \dodoi{10.1086/312840}

\bibitem[{{Green} {et~al.}(2012){Green}, {Froning}, {Osterman}, {Ebbets},
  {Heap}, {Leitherer}, {Linsky}, {Savage}, {Sembach}, {Shull}, {Siegmund},
  {Snow}, {Spencer}, {Stern}, {Stocke}, {Welsh}, {B{\'e}land}, {Burgh},
  {Danforth}, {France}, {Keeney}, {McPhate}, {Penton}, {Andrews},
  {Brownsberger}, {Morse}, \& {Wilkinson}}]{COS}
{Green}, J.~C., {Froning}, C.~S., {Osterman}, S., {et~al.} 2012, \apj, 744, 60,
  \dodoi{10.1088/0004-637X/744/1/6010.1086/141956}

\bibitem[{{Guo} {et~al.}(2018){Guo}, {Shen}, \& {Wang}}]{Pyqsofit}
{Guo}, H., {Shen}, Y., \& {Wang}, S. 2018, {PyQSOFit: Python code to fit the
  spectrum of quasars}, Astrophysics Source Code Library.
\newblock \doeprint{1809.008}

\bibitem[{{Haas} {et~al.}(2012){Haas}, {Schaye}, \&
  {Jeeson-Daniel}}]{Haas_etal_2012}
{Haas}, M.~R., {Schaye}, J., \& {Jeeson-Daniel}, A. 2012, \mnras, 419, 2133,
  \dodoi{10.1111/j.1365-2966.2011.19863.x}

\bibitem[{{Hale} {et~al.}(2021){Hale}, {McConnell}, {Thomson}, {Lenc}, {Heald},
  {Hotan}, {Leung}, {Moss}, {Murphy}, {Pritchard}, {Sadler}, {Stewart}, \&
  {Whiting}}]{Hale_etal_2021_RACS}
{Hale}, C.~L., {McConnell}, D., {Thomson}, A.~J.~M., {et~al.} 2021, arXiv
  e-prints, arXiv:2109.00956.
\newblock \doarXiv{2109.00956}

\bibitem[{{Hatch} {et~al.}(2014){Hatch}, {Wylezalek}, {Kurk}, {Stern}, {De
  Breuck}, {Jarvis}, {Galametz}, {Gonzalez}, {Hartley}, {Mortlock}, {Seymour},
  \& {Stevens}}]{Hatch_etal_2014}
{Hatch}, N.~A., {Wylezalek}, D., {Kurk}, J.~D., {et~al.} 2014, \mnras, 445,
  280, \dodoi{10.1093/mnras/stu1725}

\bibitem[{{Heckman} \& {Best}(2014)}]{Heckman_Best_2014}
{Heckman}, T.~M., \& {Best}, P.~N. 2014, \araa, 52, 589,
  \dodoi{10.1146/annurev-astro-081913-035722}

\bibitem[{{Helton} {et~al.}(2021){Helton}, {Johnson}, {Greene}, \&
  {Chen}}]{Helton_etal_2021}
{Helton}, J.~M., {Johnson}, S.~D., {Greene}, J.~E., \& {Chen}, H.-W. 2021,
  \mnras, 505, 5497, \dodoi{10.1093/mnras/stab1647}

\bibitem[{{Hewett} \& {Wild}(2010)}]{Hewett_Wild_2010}
{Hewett}, P.~C., \& {Wild}, V. 2010, \mnras, 405, 2302,
  \dodoi{10.1111/j.1365-2966.2010.16648.x}

\bibitem[{{Hickox} {et~al.}(2009){Hickox}, {Jones}, {Forman}, {Murray},
  {Kochanek}, {Eisenstein}, {Jannuzi}, {Dey}, {Brown}, {Stern}, {Eisenhardt},
  {Gorjian}, {Brodwin}, {Narayan}, {Cool}, {Kenter}, {Caldwell}, \&
  {Anderson}}]{Hickox_etal_2009}
{Hickox}, R.~C., {Jones}, C., {Forman}, W.~R., {et~al.} 2009, \apj, 696, 891,
  \dodoi{10.1088/0004-637X/696/1/891}

\bibitem[{{Hopkins} {et~al.}(2006){Hopkins}, {Hernquist}, {Cox}, {Di Matteo},
  {Robertson}, \& {Springel}}]{Hopkins_etal_2006}
{Hopkins}, P.~F., {Hernquist}, L., {Cox}, T.~J., {et~al.} 2006, \apjs, 163, 1,
  \dodoi{10.1086/499298}

\bibitem[{{Jahnke} {et~al.}(2009){Jahnke}, {Bongiorno}, {Brusa}, {Capak},
  {Cappelluti}, {Cisternas}, {Civano}, {Colbert}, {Comastri}, {Elvis},
  {Hasinger}, {Ilbert}, {Impey}, {Inskip}, {Koekemoer}, {Lilly}, {Maier},
  {Merloni}, {Riechers}, {Salvato}, {Schinnerer}, {Scoville}, {Silverman},
  {Taniguchi}, {Trump}, \& {Yan}}]{Jahnke_etal_2009}
{Jahnke}, K., {Bongiorno}, A., {Brusa}, M., {et~al.} 2009, \apjl, 706, L215,
  \dodoi{10.1088/0004-637X/706/2/L215}

\bibitem[{{Johnson} {et~al.}(2018){Johnson}, {Chen}, {Straka}, {Schaye},
  {Cantalupo}, {Wendt}, {Muzahid}, {Bouch{\'e}}, {Herenz}, {Kollatschny},
  {Mulchaey}, {Marino}, {Maseda}, \& {Wisotzki}}]{Johnson_etal_2018}
{Johnson}, S.~D., {Chen}, H.-W., {Straka}, L.~A., {et~al.} 2018, \apjl, 869,
  L1, \dodoi{10.3847/2041-8213/aaf1cf}

\bibitem[{{Karhunen} {et~al.}(2014){Karhunen}, {Kotilainen}, {Falomo}, \&
  {Bettoni}}]{Karhunen_etal_2014}
{Karhunen}, K., {Kotilainen}, J.~K., {Falomo}, R., \& {Bettoni}, D. 2014,
  \mnras, 441, 1802, \dodoi{10.1093/mnras/stu688}

\bibitem[{{Kellermann} {et~al.}(1989){Kellermann}, {Sramek}, {Schmidt},
  {Shaffer}, \& {Green}}]{Kellermann_etal_1989}
{Kellermann}, K.~I., {Sramek}, R., {Schmidt}, M., {Shaffer}, D.~B., \& {Green},
  R. 1989, \aj, 98, 1195, \dodoi{10.1086/115207}

\bibitem[{{Kelson}(2003)}]{Carpy2}
{Kelson}, D.~D. 2003, \pasp, 115, 688, \dodoi{10.1086/375502}

\bibitem[{{Kelson} {et~al.}(2000){Kelson}, {Illingworth}, {van Dokkum}, \&
  {Franx}}]{Carpy1}
{Kelson}, D.~D., {Illingworth}, G.~D., {van Dokkum}, P.~G., \& {Franx}, M.
  2000, \apj, 531, 159, \dodoi{10.1086/308445}

\bibitem[{{Kelson} {et~al.}(2014){Kelson}, {Williams}, {Dressler}, {McCarthy},
  {Shectman}, {Mulchaey}, {Villanueva}, {Crane}, \&
  {Quadri}}]{Kelson_etal_2014}
{Kelson}, D.~D., {Williams}, R.~J., {Dressler}, A., {et~al.} 2014, \apj, 783,
  110, \dodoi{10.1088/0004-637X/783/2/110}

\bibitem[{{Kormendy} \& {Ho}(2013)}]{Kormendy_Ho_2013}
{Kormendy}, J., \& {Ho}, L.~C. 2013, \araa, 51, 511,
  \dodoi{10.1146/annurev-astro-082708-101811}

\bibitem[{{Kraft} {et~al.}(1991){Kraft}, {Burrows}, \&
  {Nousek}}]{Kraft_etal_1991}
{Kraft}, R.~P., {Burrows}, D.~N., \& {Nousek}, J.~A. 1991, \apj, 374, 344,
  \dodoi{10.1086/170124}

\bibitem[{{Kravtsov} {et~al.}(2018){Kravtsov}, {Vikhlinin}, \&
  {Meshcheryakov}}]{Kravtsov_etal_2018}
{Kravtsov}, A.~V., {Vikhlinin}, A.~A., \& {Meshcheryakov}, A.~V. 2018,
  Astronomy Letters, 44, 8, \dodoi{10.1134/S1063773717120015}

\bibitem[{{Kroupa}(2001)}]{Kroupa_2001}
{Kroupa}, P. 2001, \mnras, 322, 231, \dodoi{10.1046/j.1365-8711.2001.04022.x}

\bibitem[{{Laor}(1998)}]{Laor_1998}
{Laor}, A. 1998, \apjl, 505, L83, \dodoi{10.1086/311619}

\bibitem[{{Lin} \& {Mohr}(2004)}]{Lin_Mohr_2014}
{Lin}, Y.-T., \& {Mohr}, J.~J. 2004, \apj, 617, 879, \dodoi{10.1086/425412}

\bibitem[{{Magorrian} {et~al.}(1998){Magorrian}, {Tremaine}, {Richstone},
  {Bender}, {Bower}, {Dressler}, {Faber}, {Gebhardt}, {Green}, {Grillmair},
  {Kormendy}, \& {Lauer}}]{Magorrian_etal_1998}
{Magorrian}, J., {Tremaine}, S., {Richstone}, D., {et~al.} 1998, \aj, 115,
  2285, \dodoi{10.1086/300353}

\bibitem[{{McLure} \& {Dunlop}(2001)}]{Mclure_Dunlop_2001}
{McLure}, R.~J., \& {Dunlop}, J.~S. 2001, \mnras, 321, 515,
  \dodoi{10.1046/j.1365-8711.2001.04087.x}

\bibitem[{{Munari} {et~al.}(2013){Munari}, {Biviano}, {Borgani}, {Murante}, \&
  {Fabjan}}]{Munari_etal_2013}
{Munari}, E., {Biviano}, A., {Borgani}, S., {Murante}, G., \& {Fabjan}, D.
  2013, \mnras, 430, 2638, \dodoi{10.1093/mnras/stt049}

\bibitem[{{Peng} {et~al.}(2002){Peng}, {Ho}, {Impey}, \& {Rix}}]{GALFIT}
{Peng}, C.~Y., {Ho}, L.~C., {Impey}, C.~D., \& {Rix}, H.-W. 2002, \aj, 124,
  266, \dodoi{10.1086/340952}

\bibitem[{{Peng} {et~al.}(2006){Peng}, {Impey}, {Ho}, {Barton}, \&
  {Rix}}]{Peng_etal_2006a}
{Peng}, C.~Y., {Impey}, C.~D., {Ho}, L.~C., {Barton}, E.~J., \& {Rix}, H.-W.
  2006, \apj, 640, 114, \dodoi{10.1086/499930}

\bibitem[{{Persson} {et~al.}(2013){Persson}, {Murphy}, {Smee}, {Birk},
  {Monson}, {Uomoto}, {Koch}, {Shectman}, {Barkhouser}, {Orndorff}, {Hammond},
  {Harding}, {Scharfstein}, {Kelson}, {Marshall}, \& {McCarthy}}]{Fourstar}
{Persson}, S.~E., {Murphy}, D.~C., {Smee}, S., {et~al.} 2013, \pasp, 125, 654,
  \dodoi{10.1086/671164}

\bibitem[{{Planck Collaboration} {et~al.}(2020){Planck Collaboration},
  {Aghanim}, {Akrami}, {Ashdown}, {Aumont}, {Baccigalupi}, {Ballardini},
  {Banday}, {Barreiro}, {Bartolo}, {Basak}, {Battye}, {Benabed}, {Bernard},
  {Bersanelli}, {Bielewicz}, {Bock}, {Bond}, {Borrill}, {Bouchet}, {Boulanger},
  {Bucher}, {Burigana}, {Butler}, {Calabrese}, {Cardoso}, {Carron},
  {Challinor}, {Chiang}, {Chluba}, {Colombo}, {Combet}, {Contreras}, {Crill},
  {Cuttaia}, {de Bernardis}, {de Zotti}, {Delabrouille}, {Delouis}, {Di
  Valentino}, {Diego}, {Dor{\'e}}, {Douspis}, {Ducout}, {Dupac}, {Dusini},
  {Efstathiou}, {Elsner}, {En{\ss}lin}, {Eriksen}, {Fantaye}, {Farhang},
  {Fergusson}, {Fernandez-Cobos}, {Finelli}, {Forastieri}, {Frailis},
  {Fraisse}, {Franceschi}, {Frolov}, {Galeotta}, {Galli}, {Ganga},
  {G{\'e}nova-Santos}, {Gerbino}, {Ghosh}, {Gonz{\'a}lez-Nuevo}, {G{\'o}rski},
  {Gratton}, {Gruppuso}, {Gudmundsson}, {Hamann}, {Handley}, {Hansen},
  {Herranz}, {Hildebrandt}, {Hivon}, {Huang}, {Jaffe}, {Jones}, {Karakci},
  {Keih{\"a}nen}, {Keskitalo}, {Kiiveri}, {Kim}, {Kisner}, {Knox},
  {Krachmalnicoff}, {Kunz}, {Kurki-Suonio}, {Lagache}, {Lamarre}, {Lasenby},
  {Lattanzi}, {Lawrence}, {Le Jeune}, {Lemos}, {Lesgourgues}, {Levrier},
  {Lewis}, {Liguori}, {Lilje}, {Lilley}, {Lindholm}, {L{\'o}pez-Caniego},
  {Lubin}, {Ma}, {Mac{\'\i}as-P{\'e}rez}, {Maggio}, {Maino}, {Mandolesi},
  {Mangilli}, {Marcos-Caballero}, {Maris}, {Martin}, {Martinelli},
  {Mart{\'\i}nez-Gonz{\'a}lez}, {Matarrese}, {Mauri}, {McEwen}, {Meinhold},
  {Melchiorri}, {Mennella}, {Migliaccio}, {Millea}, {Mitra},
  {Miville-Desch{\^e}nes}, {Molinari}, {Montier}, {Morgante}, {Moss}, {Natoli},
  {N{\o}rgaard-Nielsen}, {Pagano}, {Paoletti}, {Partridge}, {Patanchon},
  {Peiris}, {Perrotta}, {Pettorino}, {Piacentini}, {Polastri}, {Polenta},
  {Puget}, {Rachen}, {Reinecke}, {Remazeilles}, {Renzi}, {Rocha}, {Rosset},
  {Roudier}, {Rubi{\~n}o-Mart{\'\i}n}, {Ruiz-Granados}, {Salvati}, {Sandri},
  {Savelainen}, {Scott}, {Shellard}, {Sirignano}, {Sirri}, {Spencer},
  {Sunyaev}, {Suur-Uski}, {Tauber}, {Tavagnacco}, {Tenti}, {Toffolatti},
  {Tomasi}, {Trombetti}, {Valenziano}, {Valiviita}, {Van Tent}, {Vibert},
  {Vielva}, {Villa}, {Vittorio}, {Wandelt}, {Wehus}, {White}, {White},
  {Zacchei}, \& {Zonca}}]{Planck_2018}
{Planck Collaboration}, {Aghanim}, N., {Akrami}, Y., {et~al.} 2020, \aap, 641,
  A6, \dodoi{10.1051/0004-6361/201833910}

\bibitem[{{Qu} {et~al.}(2023){Qu}, {Chen}, {Rudie}, {Johnson}, {Zahedy},
  {DePalma}, {Boettcher}, {Cantalupo}, {Chen}, {Cooksey},
  {Faucher-Gigu{\`e}re}, {Li}, {Lopez}, {Schaye}, \& {Simcoe}}]{Qu_etal_2023}
{Qu}, Z., {Chen}, H.-W., {Rudie}, G.~C., {et~al.} 2023, \mnras, 524, 512,
  \dodoi{10.1093/mnras/stad1886}

\bibitem[{{Ramos Almeida} {et~al.}(2013){Ramos Almeida}, {Bessiere},
  {Tadhunter}, {Inskip}, {Morganti}, {Dicken}, {Gonz{\'a}lez-Serrano}, \&
  {Holt}}]{RamosAlmeida_etal_2013}
{Ramos Almeida}, C., {Bessiere}, P.~S., {Tadhunter}, C.~N., {et~al.} 2013,
  \mnras, 436, 997, \dodoi{10.1093/mnras/stt1595}

\bibitem[{{Richardson} {et~al.}(2012){Richardson}, {Zheng}, {Chatterjee},
  {Nagai}, \& {Shen}}]{Richardson_etal_2012}
{Richardson}, J., {Zheng}, Z., {Chatterjee}, S., {Nagai}, D., \& {Shen}, Y.
  2012, \apj, 755, 30, \dodoi{10.1088/0004-637X/755/1/30}

\bibitem[{{Rudie} {et~al.}(2017){Rudie}, {Newman}, \&
  {Murphy}}]{Rudie_etal_2017}
{Rudie}, G.~C., {Newman}, A.~B., \& {Murphy}, M.~T. 2017, \apj, 843, 98,
  \dodoi{10.3847/1538-4357/aa74d7}

\bibitem[{{Serber} {et~al.}(2006){Serber}, {Bahcall}, {M{\'e}nard}, \&
  {Richards}}]{Serber_etal_2006}
{Serber}, W., {Bahcall}, N., {M{\'e}nard}, B., \& {Richards}, G. 2006, \apj,
  643, 68, \dodoi{10.1086/501443}

\bibitem[{{Sevilla-Noarbe} {et~al.}(2021){Sevilla-Noarbe}, {Bechtol}, {Carrasco
  Kind}, {Carnero Rosell}, {Becker}, {Drlica-Wagner}, {Gruendl}, {Rykoff},
  {Sheldon}, {Yanny}, {Alarcon}, {Allam}, {Amon}, {Benoit-L{\'e}vy},
  {Bernstein}, {Bertin}, {Burke}, {Carretero}, {Choi}, {Diehl}, {Everett},
  {Flaugher}, {Gaztanaga}, {Gschwend}, {Harrison}, {Hartley}, {Hoyle},
  {Jarvis}, {Johnson}, {Kessler}, {Kron}, {Kuropatkin}, {Leistedt}, {Li},
  {Menanteau}, {Morganson}, {Ogando}, {Palmese}, {Paz-Chinch{\'o}n}, {Pieres},
  {Pond}, {Rodriguez-Monroy}, {Smith}, {Stringer}, {Troxel}, {Tucker}, {de
  Vicente}, {Wester}, {Zhang}, {Abbott}, {Aguena}, {Annis}, {Avila},
  {Bhargava}, {Bridle}, {Brooks}, {Brout}, {Castander}, {Cawthon}, {Chang},
  {Conselice}, {Costanzi}, {Crocce}, {da Costa}, {Pereira}, {Davis}, {Desai},
  {Dietrich}, {Doel}, {Eckert}, {Evrard}, {Ferrero}, {Fosalba},
  {Garc{\'\i}a-Bellido}, {Gerdes}, {Giannantonio}, {Gruen}, {Gutierrez},
  {Hinton}, {Hollowood}, {Honscheid}, {Huff}, {Huterer}, {James}, {Jeltema},
  {Kuehn}, {Lahav}, {Lidman}, {Lima}, {Lin}, {Maia}, {Marshall}, {Martini},
  {Melchior}, {Miquel}, {Mohr}, {Morgan}, {Neilsen}, {Plazas}, {Romer},
  {Roodman}, {Sanchez}, {Scarpine}, {Schubnell}, {Serrano}, {Smith}, {Suchyta},
  {Tarle}, {Thomas}, {To}, {Varga}, {Wechsler}, {Weller}, {Wilkinson}, \& {DES
  Collaboration}}]{DES_Y3}
{Sevilla-Noarbe}, I., {Bechtol}, K., {Carrasco Kind}, M., {et~al.} 2021, \apjs,
  254, 24, \dodoi{10.3847/1538-4365/abeb66}

\bibitem[{{Shen} {et~al.}(2009){Shen}, {Strauss}, {Ross}, {Hall}, {Lin},
  {Richards}, {Schneider}, {Weinberg}, {Connolly}, {Fan}, {Hennawi}, {Shankar},
  {Vanden Berk}, {Bahcall}, \& {Brunner}}]{Shen_etal_2009}
{Shen}, Y., {Strauss}, M.~A., {Ross}, N.~P., {et~al.} 2009, \apj, 697, 1656,
  \dodoi{10.1088/0004-637X/697/2/1656}

\bibitem[{{Shen} {et~al.}(2011){Shen}, {Richards}, {Strauss}, {Hall},
  {Schneider}, {Snedden}, {Bizyaev}, {Brewington}, {Malanushenko},
  {Malanushenko}, {Oravetz}, {Pan}, \& {Simmons}}]{Shen_etal_2011_sdssq}
{Shen}, Y., {Richards}, G.~T., {Strauss}, M.~A., {et~al.} 2011, \apjs, 194, 45,
  \dodoi{10.1088/0067-0049/194/2/45}

\bibitem[{{Shen} {et~al.}(2013){Shen}, {McBride}, {White}, {Zheng}, {Myers},
  {Guo}, {Kirkpatrick}, {Padmanabhan}, {Parejko}, {Ross}, {Schlegel},
  {Schneider}, {Streblyanska}, {Swanson}, {Zehavi}, {Pan}, {Bizyaev},
  {Brewington}, {Ebelke}, {Malanushenko}, {Malanushenko}, {Oravetz}, {Simmons},
  \& {Snedden}}]{Shen_etal_2013}
{Shen}, Y., {McBride}, C.~K., {White}, M., {et~al.} 2013, \apj, 778, 98,
  \dodoi{10.1088/0004-637X/778/2/98}

\bibitem[{{Shen} {et~al.}(2023){Shen}, {Grier}, {Horne}, {Stone}, {Li}, {Yang},
  {Homayouni}, {Trump}, {Anderson}, {Brandt}, {Hall}, {Ho}, {Jiang},
  {Petitjean}, {Schneider}, {Tao}, {Donnan}, {AlSayyad}, {Bershady}, {Blanton},
  {Bizyaev}, {Bundy}, {Chen}, {Davis}, {Dawson}, {Fan}, {Greene}, {Groller},
  {Guo}, {Ibarra-Medel}, {Keenan}, {Kollmeier}, {Lejoly}, {Li}, {de la
  Macorra}, {Moe}, {Nie}, {Rossi}, {Smith}, {Tee}, {Weijmans}, {Xu}, {Yue},
  {Zhou}, {Zhou}, \& {Zou}}]{shen_etal_2023}
{Shen}, Y., {Grier}, C.~J., {Horne}, K., {et~al.} 2023, arXiv e-prints,
  arXiv:2305.01014, \dodoi{10.48550/arXiv.2305.01014}

\bibitem[{{Silk} \& {Rees}(1998)}]{Silk_Rees_1998}
{Silk}, J., \& {Rees}, M.~J. 1998, \aap, 331, L1.
\newblock \doarXiv{astro-ph/9801013}

\bibitem[{{Stocke} {et~al.}(1992){Stocke}, {Morris}, {Weymann}, \&
  {Foltz}}]{Stocke_etal_1992}
{Stocke}, J.~T., {Morris}, S.~L., {Weymann}, R.~J., \& {Foltz}, C.~B. 1992,
  \apj, 396, 487, \dodoi{10.1086/171735}

\bibitem[{{Stone} {et~al.}(2023){Stone}, {Wethers}, {de Propris}, {Kotilainen},
  {Acharya}, {Holwerda}, {Loveday}, \& {Phillipps}}]{Stone_etal_2023}
{Stone}, M.~B., {Wethers}, C.~F., {de Propris}, R., {et~al.} 2023, \apj, 946,
  116, \dodoi{10.3847/1538-4357/acbd4d}

\bibitem[{{Stott} {et~al.}(2020){Stott}, {Bielby}, {Cullen}, {Burchett},
  {Tejos}, {Fumagalli}, {Crain}, {Morris}, {Amos}, {Bower}, \&
  {Prochaska}}]{Stott_etal_2020}
{Stott}, J.~P., {Bielby}, R.~M., {Cullen}, F., {et~al.} 2020, \mnras, 497,
  3083, \dodoi{10.1093/mnras/staa2096}

\bibitem[{{Tinker} {et~al.}(2013){Tinker}, {Leauthaud}, {Bundy}, {George},
  {Behroozi}, {Massey}, {Rhodes}, \& {Wechsler}}]{Tinker_etal_2013}
{Tinker}, J.~L., {Leauthaud}, A., {Bundy}, K., {et~al.} 2013, \apj, 778, 93,
  \dodoi{10.1088/0004-637X/778/2/93}

\bibitem[{{Trainor} \& {Steidel}(2012)}]{Trainor_etal_2012}
{Trainor}, R.~F., \& {Steidel}, C.~C. 2012, \apj, 752, 39,
  \dodoi{10.1088/0004-637X/752/1/39}

\bibitem[{{Treu} {et~al.}(2004){Treu}, {Malkan}, \&
  {Blandford}}]{Treu_etal_2004}
{Treu}, T., {Malkan}, M.~A., \& {Blandford}, R.~D. 2004, \apjl, 615, L97,
  \dodoi{10.1086/426437}

\bibitem[{{Voit} {et~al.}(2023){Voit}, {Oppenheimer}, {Bell}, {Terrazas}, \&
  {Donahue}}]{Voit_etal_2023}
{Voit}, G.~M., {Oppenheimer}, B.~D., {Bell}, E.~F., {Terrazas}, B., \&
  {Donahue}, M. 2023, arXiv e-prints, arXiv:2309.14818,
  \dodoi{10.48550/arXiv.2309.14818}

\bibitem[{{Wechsler} \& {Tinker}(2018)}]{Wechsler_Tinker_2018}
{Wechsler}, R.~H., \& {Tinker}, J.~L. 2018, \araa, 56, 435,
  \dodoi{10.1146/annurev-astro-081817-051756}

\bibitem[{{Weilbacher} {et~al.}(2020){Weilbacher}, {Palsa}, {Streicher},
  {Bacon}, {Urrutia}, {Wisotzki}, {Conseil}, {Husemann}, {Jarno}, {Kelz},
  {P{\'e}contal-Rousset}, {Richard}, {Roth}, {Selman}, \&
  {Vernet}}]{Weilbacher_etal_2020}
{Weilbacher}, P.~M., {Palsa}, R., {Streicher}, O., {et~al.} 2020, \aap, 641,
  A28, \dodoi{10.1051/0004-6361/202037855}

\bibitem[{{Wethers} {et~al.}(2022){Wethers}, {Acharya}, {De Propris},
  {Kotilainen}, {Baldry}, {Brough}, {Driver}, {Graham}, {Holwerda}, {Hopkins},
  {L{\'o}pez-S{\'a}nchez}, {Loveday}, {Phillipps}, {Pimbblet}, {Taylor},
  {Wang}, \& {Wright}}]{Wethers_etal_2022}
{Wethers}, C.~F., {Acharya}, N., {De Propris}, R., {et~al.} 2022, \apj, 928,
  192, \dodoi{10.3847/1538-4357/ac409c}

\bibitem[{{Woo} {et~al.}(2006){Woo}, {Treu}, {Malkan}, \&
  {Blandford}}]{Woo_etal_2006}
{Woo}, J.-H., {Treu}, T., {Malkan}, M.~A., \& {Blandford}, R.~D. 2006, \apj,
  645, 900, \dodoi{10.1086/504586}

\bibitem[{{Woo} {et~al.}(2010){Woo}, {Treu}, {Barth}, {Wright}, {Walsh},
  {Bentz}, {Martini}, {Bennert}, {Canalizo}, {Filippenko}, {Gates}, {Greene},
  {Li}, {Malkan}, {Stern}, \& {Minezaki}}]{Woo_etal_2010}
{Woo}, J.-H., {Treu}, T., {Barth}, A.~J., {et~al.} 2010, \apj, 716, 269,
  \dodoi{10.1088/0004-637X/716/1/269}

\bibitem[{{Wylezalek} {et~al.}(2013){Wylezalek}, {Galametz}, {Stern}, {Vernet},
  {De Breuck}, {Seymour}, {Brodwin}, {Eisenhardt}, {Gonzalez}, {Hatch},
  {Jarvis}, {Rettura}, {Stanford}, \& {Stevens}}]{Wylezalek_etal_2013}
{Wylezalek}, D., {Galametz}, A., {Stern}, D., {et~al.} 2013, \apj, 769, 79,
  \dodoi{10.1088/0004-637X/769/1/79}

\bibitem[{{Zhang} {et~al.}(2013){Zhang}, {Wang}, {Wang}, \&
  {Zhou}}]{Zhang_etal_2013}
{Zhang}, S., {Wang}, T., {Wang}, H., \& {Zhou}, H. 2013, \apj, 773, 175,
  \dodoi{10.1088/0004-637X/773/2/175}

\end{thebibliography}

\end{document}